\documentclass[a4paper,11pt]{article}
\pdfoutput=1
\usepackage[T1]{fontenc} 
\usepackage{booktabs} 
\usepackage{algorithm} 
\usepackage{xfrac} 
\usepackage{graphicx}
\usepackage{caption}
\usepackage{subcaption}
\usepackage{jcappub}
\usepackage{orcidlink} 
\usepackage{adjustbox} 
\usepackage{pdflscape} 
\usepackage{soul} 
\usepackage{tikz} 
\usepackage{comment}
\usepackage{float} 
\usepackage{amssymb}   
\usepackage{amsmath}   
\usepackage{bm}
\usepackage{enumitem}
\newlist{enumerate*}{enumerate}{1}
\setlist[enumerate*]{label=(\roman*), before*=\unskip{}, itemjoin={{; }}, itemjoin*={{; and }}}

\usepackage{xspace}
\usepackage{verbatim}
\usepackage{tikz}
\renewcommand{\d}{\mathrm{d}}
\usepackage{babel} 
\usepackage{multicol} 
\usepackage[T1]{fontenc}
\usepackage[utf8]{inputenc}
\usepackage{babel}
\usepackage[font=small,labelfont=bf]{caption}
\usepackage{enumitem}
\usepackage{makecell} 
\usepackage{multirow} 
\usepackage[T1]{fontenc}
\usepackage{aecompl}
\usepackage{makecell}
\usepackage{graphicx}
\usepackage{rotating,tabularx} 
\usepackage{float} 
\usepackage{colortbl} 
\usepackage{xcolor} 
\definecolor{LightGray}{gray}{0.8}
\definecolor{LighterGray}{gray}{0.9}
\definecolor{mygreen}{rgb}{0,0.5,0}
\definecolor{brightred}{rgb}{1,0.4,0.4}
\definecolor{brightblue}{rgb}{0.4,0.4,1}

\newcommand{\X}[1]{\ensuremath{a_{#1}\! /\! a}}
\newcommand{\Y}[1]{\ensuremath{\phi_{#1}\!-\!\phi_0}}
\newcommand{\Z}{\ensuremath{q}}

\newcommand{\farcs}{\mbox{$.\!\!^{\prime\prime}$}}

\title{%
Joint Semi-Analytic Multipole Priors from Galaxy Isophotes and Constraints from Lensed Arcs
}

\def\be{\begin{equation}}
\def\ee{\end{equation}}

\author[a,1]{Maverick S. H. Oh \orcidlink{0000-0003-0772-4100},\note{Corresponding author.}}

\affiliation[a]{University of California, Merced,\\5200 North Lake Rd.
Merced, CA 95343, USA}

\author[a]{Anna Nierenberg \orcidlink{0000-0001-6809-2536},}
\author[b,c]{Daniel Gilman \orcidlink{0000-0002-5116-7287},}
\affiliation[b]{The University of Chicago,\\5801 S. Ellis Ave.
Chicago, IL 60637, USA}
\affiliation[c]{Brinson Prize Fellow}

\author[d]{Simon Birrer \orcidlink{0000-0003-3195-5507}}
\affiliation[d]{Stony Brook University,\\100 Nicolls Rd, Stony Brook, NY 11794, USA}

\emailAdd{soh39@ucmerced.edu}

\abstract{Flux-ratio anomalies in quadruply imaged quasars are sensitive to the imprint of low-mass
dark-matter haloes. The reliability of detection depends on the robustness of the smooth mass model. Optical surveys show that massive early-type galaxies similar to galaxy-scale gravitational lenses
depart from perfect ellipticity, exhibiting $m=3$ and $m=4$
multipole distortions. 
We construct the semi-analytic, five-dimensional joint population prior for the $m=3$ and $m=4$ amplitude and orientation as well as the axis ratio of the deflector, calibrated on the sample of 840 SDSS E/S0 galaxies. The parameters are fitted via hierarchical
Bayesian modeling, minimizing a joint Jensen–Shannon divergence
between model and data. We use this prior to model the mass distribution of mock lenses with HST quality data with different multipole amplitudes. We find that we robustly measure the true multipole amplitudes and orientations. Compared to fits that use only the four point-image positions, adding the lensed host-galaxy
arcs tightens the 68 \% credible regions of multipole parameters by factors of $3$–$12$ and reduces the predicted flux-ratio
uncertainties by a mean factor of~${\sim}6$. 
This analysis does not include substructure or a complex source, and thus can be considered an upper limit on the expected improvement.
The combination of
arc information and realistic multipole priors therefore yields an
order-of-magnitude improvement in smooth mass model precision, paving the way for
more robust measurements of dark-matter substructure.}

\begin{document}
\maketitle
\flushbottom

\setcounter{footnote}{1}


\section{Introduction}\label{sec:intro}

Dark matter dominates the mass budget of the Universe, yet its particle nature remains unknown \cite{ade2014planck}.
Strong gravitational lensing of quasars provides one of the few probes of dark matter on sub-galactic scales because the deflection angles are set directly by the mass distribution—even in the absence of luminous tracers \citep{vegetti2023strong}. While various methods such as stellar stream perturbations and galaxy kinematics use baryonic tracers to infer the presence of low-mass dark matter halos \citep[e.g.][]{Bonaca_2019,Simon_2019}, these approaches are often subject to degeneracies with baryonic physics. In contrast, gravitational lensing offers a more direct and clean probe of the mass distribution, enabling discrimination among dark matter models based on the abundance and internal structure of subhalos \citep[see][and references therein]{vegetti2023strong}.

The flux ratios of multiple quasar images are among the key observables of strong gravitational lensing. Flux ratio anomalies refer to the disparity between the observed flux ratios of lensed images and the flux ratios expected from a smooth mass distribution, which represents the large-scale distribution of the mass of the main 
lensing galaxy and its dark matter halo
(a.k.a. ``macromodel''). 
The macromodel is primarily responsible for causing multiple images of the background source to appear and determining their positions. 
Low-mass dark matter halos within the lensing galaxy and along the line of sight (a.k.a. ``substructure'') make relatively little impact to the image positions relative to measurement uncertainties, but can introduce perturbations to the lensed image magnifications \citep[see][and references therein]{mao1998evidence,Dalal_2002,Kochanek_2004,vegetti2023strong}.
Note that such flux ratio anomalies may also arise from microlensing \citep[e.g.][]{Schechter_2002} or inaccuracies in the smooth lens model \citep[e.g.][]{xu2015well}. These effects need to be considered carefully when attributing flux ratio anomalies to dark matter substructure.
Studies on multiple lens systems have shown that the amplitude and frequency of flux ratio anomalies exceed what can be accounted for by these effects alone, lending support to the presence of low-mass dark matter halos \citep[e.g.][]{Dalal_2002,Nierenberg2014, congdon2005multipole}.

Flux ratio anomalies can reveal the presence of low-mass substructures and have been used to infer population level statistics of their mass function and mass distribution \citep{hsueh_2019,Gilman2020,Gilman2019-1}.   
However, robust inference from flux ratios requires an accurate smooth mass model.  
Traditional analyses adopt an Elliptical Power Law (EPL) profile plus external shear \citep[e.g.][]{nierenberg2017,shajib2019every,Oriodan2019,schmidt2023strides}, while recent studies \citep[e.g.][]{hsueh_2016,Gilman++24,stacey2024complex} have begun investigating the effect of deviations from ellipticity as detected in optical surveys of lens-like elliptical galaxies
\citep[][Hao et al. 2006, hereafter H06]{hao2006isophotal}.  

Multipoles describe higher order perturbations to the mass distribution that cannot be captured by an elliptical profile  \citep{kochanek1991implications,trotter2000multipole,moller2003discs,evans2003fitting,Kochanek_2004,congdon2005multipole,gilman2017strong,Vyvere2022,stacey2024complex}. Of particular interest are multipoles of order $m\!=\!3$ and $m\!=\!4$ as these are prominent deviations from ellipticity observed in the light distribution of field elliptical galaxies \citep{hao2006isophotal}. However, \textit{mass} multipoles have not been well measured compared to \textit{light} multipoles, and there has not been empirically motivated priors on the multipole amplitudes $(\X3,\X4)$ and orientations $(\Y3,\Y4)$ as a readily usable form so far.
Without such priors, estimating the correct multipole parameters from observed data can be challenging due to degeneracy.
\citep{etherington2023strong} pointed out that, when multipoles are not included in the lens modeling, they may appear in the external shear that does not agree with an independently measured cosmic shear.

Existing constraints on dark matter from \cite{Gilman++21,Gilman++22a,Gilman++22b,Gilman++23} include an $m=4$ multipole term, which adds boxyness and diskyness to the main deflector mass profile. The amplitude of this mass component is constrained by image positions and flux ratios jointly with the substructure properties. A recent simulation work presented the dark matter analysis pipeline with both $m=3$ and $m=4$ multipoles \cite{Gilman++24}. A warm dark matter constraint using JWST MIRI observations with both $m=3$ and $m=4$ multipoles and flexible $\phi_3$, $\phi_4$ was also presented recently \cite{keeley2024jwst}. These works adopted simple Gaussian priors for the multipole distribution. We build on these works by increasing the complexity of the prior to be both more flexible, as well as conditionally dependent on ellipticity. This work provides a continuous and differentiable joint distribution of multipole parameters to be used as a realistic prior.

Extended arcs--the lensed host-galaxy light--offer a route to break the difficulty of measuring multipoles. Because arcs wrap around the deflector, they probe the lensing potential over a much larger azimuthal range than the four point images alone \citep{shajib2020strides,schmidt2023strides}. 
In this paper, we demonstrate that incorporating arc information, together with a population prior derived from H06, tightens the posterior on multipole parameters and predicted flux ratios.

Traditionally, only the lensed point source positions have been used to constrain the smooth-mass distribution. This yields large uncertainties of order 10-50\% in the underlying smooth-model flux ratios which are comparable to or larger than the measurement uncertainties of the actual flux ratios themselves \citep[see, e.g.][]{Nierenberg2019}. 
Existing measurements of lens flux ratios from HST reach, on average, 6\% precision \citep{Nierenberg2019}, while mid-IR flux ratios measured with JWST can reach precisions of 1\% \citep{nierenberg2023jwst}. Improving the precision of the smooth-model flux ratio predictions would therefore make a significant impact on the constraining power of gravitational lenses.
\par

The remainder of this paper is organized as follows.  
Section~\ref{sec:models} introduces our mass- and light-model parameterization, including the multipoles.  
Section~\ref{sec:prior} derives the population prior from the H06 catalog.  
Section~\ref{sec:mock} describes the mock–data generation and observational setups.  
Our inference framework is detailed in Section~\ref{sec:inference}, the results are presented in Section~\ref{sec:results}, and Section~\ref{sec:discussion} discusses implications for future dark-matter studies.  
Throughout we adopt $H_0 = 70\;\mathrm{km\,s^{-1}\,Mpc^{-1}}$, $\Omega_\mathrm{m}=0.3$, and $\Omega_\Lambda=0.7$. \textsc{lenstronomy}\footnote{\url{https://github.com/lenstronomy/lenstronomy}} \citep{birrer2018lenstronomy,birrer2021lenstronomy} is used for lens modeling and fitting processes. The multipole prior is released as open-source Python software \textsc{multipoleprior}\footnote{\url{https://github.com/Maverick-Oh/multipoleprior}}, along with additional scripts used for its development and visualization for transparency and reproducibility.

\section{Quadruply lensed quasar modeling}
\label{sec:models}
In this section, we discribe the model we use to describe quadruply lensed quasars and their host galaxies.
\subsection{Mass model}
\label{subsec:mass}

The mass model includes three components: elliptical base mass model, multipole perturbations, and shear. Each of them are detailed below. Note that dark matter substructures are not included in this work to clearly demonstrate the impact of multipoles and lensed arcs without substructure lensing.

\subsubsection{Elliptical base mass model}
We adopt Elliptical power-law (EPL), of which the projected surface–mass density takes the form
$
  \kappa(x,y)
  = \frac{3-\gamma}{2} \left( 
  \frac{\theta_E}{\sqrt{q x^2 + y^2\!/\!q}}
  \right)^{\gamma-1}
$
when its major axis is aligned with the $x$-axis,
where $\gamma$ is the 3-D logarithmic slope ($2$ for isothermal), $\theta_E$ the circularized Einstein radius, and
$q\!=\!b/a$ the projected axis ratio
\citep{tessore2015elliptical}. In general, the orientation of the ellipse is denoted by $\phi_0$, the angle of the major axis and $x$-axis.

\subsubsection{Multipole perturbations}
\label{subsec:multipoles}
Multipole mass profiles add azimuthal perturbation to the mass profile of the lens, on top of the elliptical profile. We use the circular multipole convention \citep{evans2003fitting, xu2015well} to be consistent with H06. However, it should be noted that the elliptical multipoles have been shown to produce more physically meaningful optical and mass distributions at high ellipticity \citep[see][]{Ciambur_2015, paugnat2025ellipticalmultipoles} and are therefore recommended for future optical surveys.

When an order-$m$ multipole is added to an elliptical profile, the isodensity (i.e. constant convergence $\kappa$) contours are deformed from a purely elliptical isodensity contours. The deviation of the new isodensity contour from the ellipse can be expressed with a cosine function
$
\delta r = a_m \cos(m (\phi-\phi_m))
$, where $r$ is the polar radius $r=\sqrt{x^2+y^2}$
and
$\phi_m$ is the angle of the multipole profile's orientation $\phi_m$. Note that the deviation amplitude $a_m$ depends on which contour is chosen. When geometric similarity is assumed between the contours on different scales, the radial deviation is proportional to the size of the ellipse; i.e. $a_m \propto a$, where $a$ is the semi-major axis of the ellipse and satisfies $a = \frac{1}{2\kappa} {\theta_E}/\!{\sqrt{q}}$. The ratio  $a_m\!/\!a$ is the key parameter that determines the shape of the deformed isodensity contour.
\par

Choosing the standard isodensity contour with 
\(
\kappa\!=\!\frac{1}{2}
\)
gives the semi-major axis of the ellipse as the effective Einstein radius 
\(
a\mbox{\footnotesize $(\kappa\!=\!\frac{1}{2})$ } = \theta_E/\!\sqrt{q}
\).
The deviation amplitude there, 
\(
a_m\mbox{\footnotesize $(\kappa\!=\!\frac{1}{2})$}
\)
, is used for setting the deflection potential. The deviation of the isodensity contours from the standard ellipse with $\kappa\!=\!\frac{1}{2}$ is 
\begin{equation}
\delta r \rvert_{\kappa=\frac{1}{2}} = a_{m}\mbox{\footnotesize $(\kappa\!=\!\frac{1}{2})$} \ \cos(m (\phi-\phi_m)).
\end{equation}
With a given value of $a_m\!/\!a$, the $a_m \scriptstyle (\kappa=\frac{1}{2})$ value is calculated by
\begin{equation}
    a_m\mbox{\footnotesize $(\kappa\!=\!\frac{1}{2})$} = 
a_m\!/\!a \ 
a\mbox{\mbox{\footnotesize $(\kappa\!=\!\frac{1}{2})$}}
=
\ a_m\!/\!a \ \frac{\theta_E}{\sqrt{q}}.
\end{equation}
Thus, the deflection potential can be set up based on either $a_m\!/\!a$ or $a_m\mbox{\footnotesize $(\kappa\!=\!\frac{1}{2})$}$. In this paper, we stick to $a_m\!/\!a$ convention.
\par

\begin{figure*}[]
\centering
    \includegraphics[width=\textwidth]{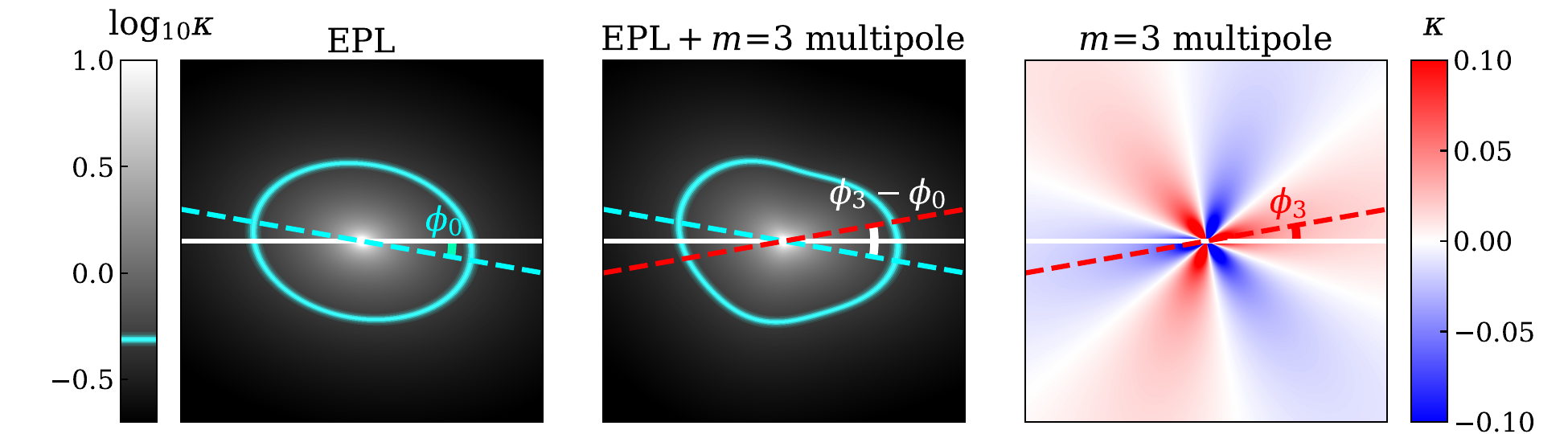}
\includegraphics[width=\textwidth]{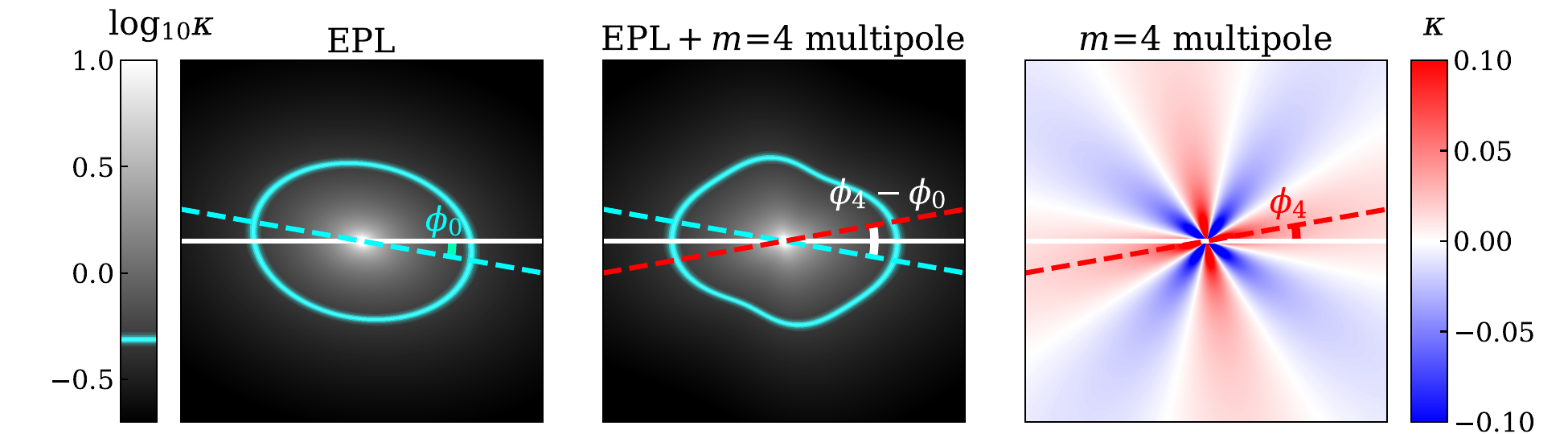}
\caption{
Impact of \textit{misaligned} ($\phi_m \!\neq\! \phi_0$) $m\!=\!3$ \textbf{(upper)} and $m\!=\!4$ \textbf{(lower)} multipole profile on the convergence with the angular conventions $\phi_0$ and $\phi_m$. The isodensity curve is shown in cyan.
 \textbf{(Left)} Convergence of EPL-only lens mass model with $\phi_0 = -0.175 \ (-10^\circ)$.
 \textbf{(Middle)} Convergence of EPL+multipole. 
 \textbf{(Right)} Convergence of $m\!=\!3$ and $m\!=\!4$ multipole with $a_m\!/\!a=0.05$ and $\phi_m = 0.175 \ (10^\circ)$.
 For better visualization, $|a_m/a|$ is set to be larger than expected in typical systems.
 }
    \label{fig:boxydisky_misaligned}
\end{figure*}

\begin{figure*}[]
 \centering
 \includegraphics[scale=0.4]{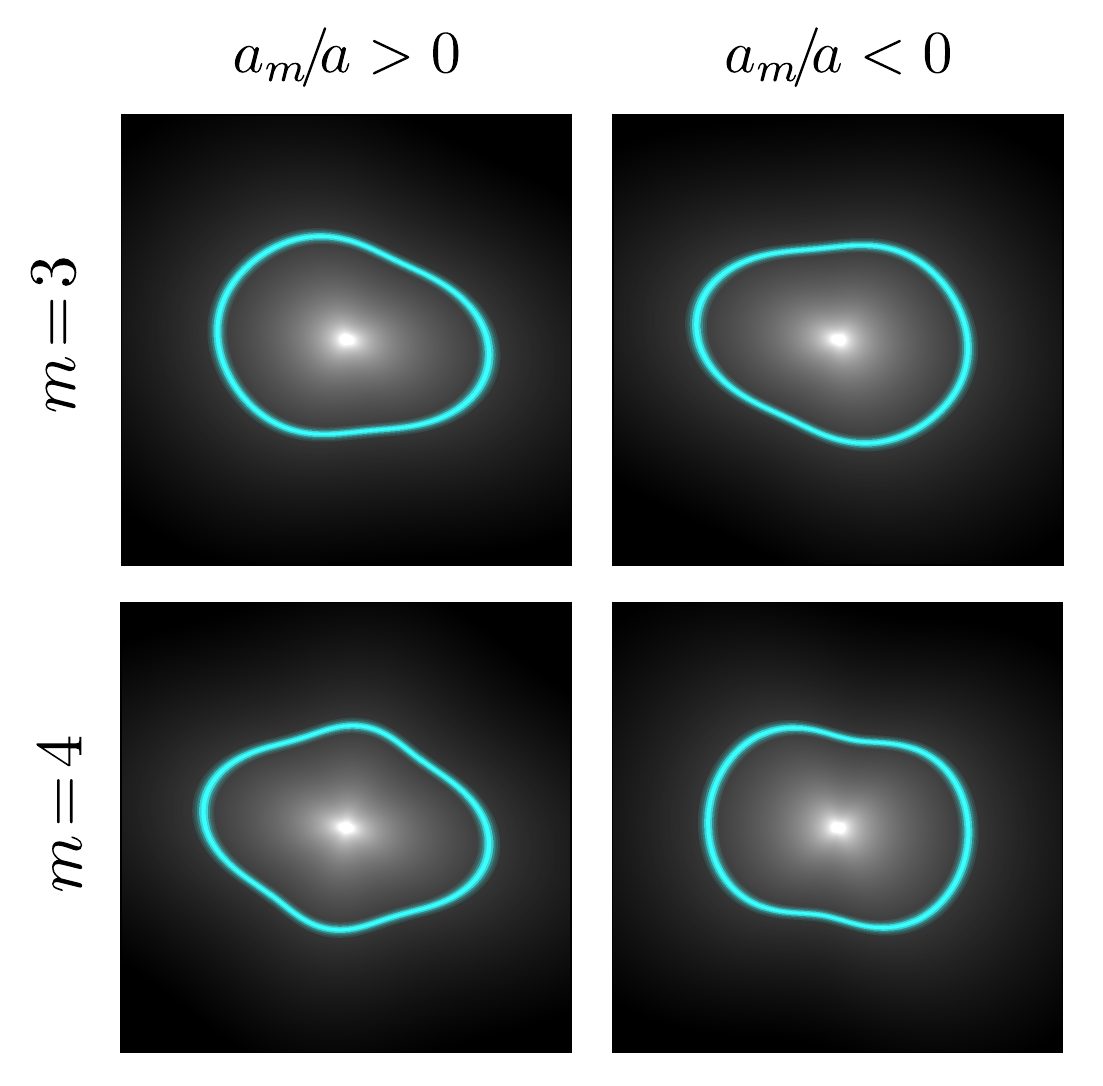}
 \caption{
 Impact of \textit{aligned} ($\phi_m \!=\! \phi_0$) $m\!=\!3$ (upper) and $m\!=\!4$ (lower) multipole profile on the convergence with different signs of $a_m\!/\!a$.
 \textbf{(Left)} Convergence of EPL+multipole with $a_m\!/\!a=+0.05$.
 \textbf{(Right)} Convergence of EPL+multipole with $a_m\!/\!a=-0.05$.
 For better visualization, $|a_m\!/\!a|$ is set to be larger than expected in typical systems. An isodensity contour is highlighted in cyan color.
 }
\label{fig:boxydisky}
\end{figure*}

We are interested in studying $m=3$ and $m=4$ multipoles.
The $m\!=\!3$ and $m\!=\!4$ multipoles measure the triangle-like and quadrangle-like deformation on the isodensity contour, respectively. They are parameterized with their multipole strength $\X{m}$ and their angle relative to the elliptical profile $\Y{m}$, where $\phi_0$ refers to the angle of the elliptical profile (see Figure \ref{fig:boxydisky_misaligned}).
\par

The $m\!=\!4$ multipole is also known as boxy/diskyness, because it makes an elliptical profile either boxy or disky. When the orientation of the multipole profile and the elliptical profile are well aligned (i.e. $\phi_m\approx\phi_0$, where $\phi_0$ is the angle of the elliptical profile), $a_4\!/\!a > 0$ results in a disky profile and $a_4\!/\!a < 0$ a boxy profile. Meanwhile, the sign of $a_3\!/\!a$ changes only the orientation of the deformation, not its overall shape (see Figure \ref{fig:boxydisky}).

Note that there are different conventions on how to define multipole variables and we follow the convention of \cite{Keeton_2003,xu2015well,Vyvere2022}. 
We allow both signs of $a_m$ and restrict
$\phi_m\in(-\pi/2m,\pi/2m]$ to avoid angular degeneracy.
In Appendix \ref{appendix:multipole_convention} we provide relations between this and other commonly used conventions.

\subsubsection{External shear}
We include external shear which can be caused by external sources such as galaxy clusters or large-scale structure
\cite{schneider1994steps,schneider2006gravitational,meneghetti2021introduction}.

\subsection{Light components}
\label{subsec:light}

The deflector light and the quasar-host arcs are each represented by
an elliptical Sérsic profile \citep{sersic1963influence}. Deflected quasar images are treated as point sources
(delta functions convolved with the PSF).


\section{Population prior for multipole parameter values}
\label{sec:prior}

In this section we explore what reasonable priors are for the mass multipole parameters.
The multipole parameters of the lens mass profiles have not been well constrained from the observation so far. 
However, \textit{optical} surveys of elliptical-galaxy light profiles can serve as empirical benchmarks and place conservative upper limits on multipole amplitudes, because
(i) dark-matter halos are well described by simple ellipsoids with negligible higher-order structure \citep[see][and references therein]{Allgood2006shapeof,Despali2014somelikeit};
(ii) the baryonic disc and bulge introduce non-elliptical perturbations \citep[e.g.][]{hsueh_2016}; and
(iii) baryons contribute only $\sim30$–$40\,\%$ of the projected mass within the Einstein radius \citep{Auger_2009,Barnabe2011}.
The optical $a_m$ and $a$ are defined the similarly as the mass profile's isodensity contours, but instead using isophote contours. The ratio of deformation amplitude and the semi-major axis, $a_m\!/\!a$, has been measured with different galaxies and at different radii 
\citep{rest2001wfpc2,hao2006isophotal,Pasquali_2006,Mitsuda_2017,stacey2024complex,he2024unveiling}.
\par

\begin{figure}[H]
    \centering
    \includegraphics[width=0.85\linewidth]{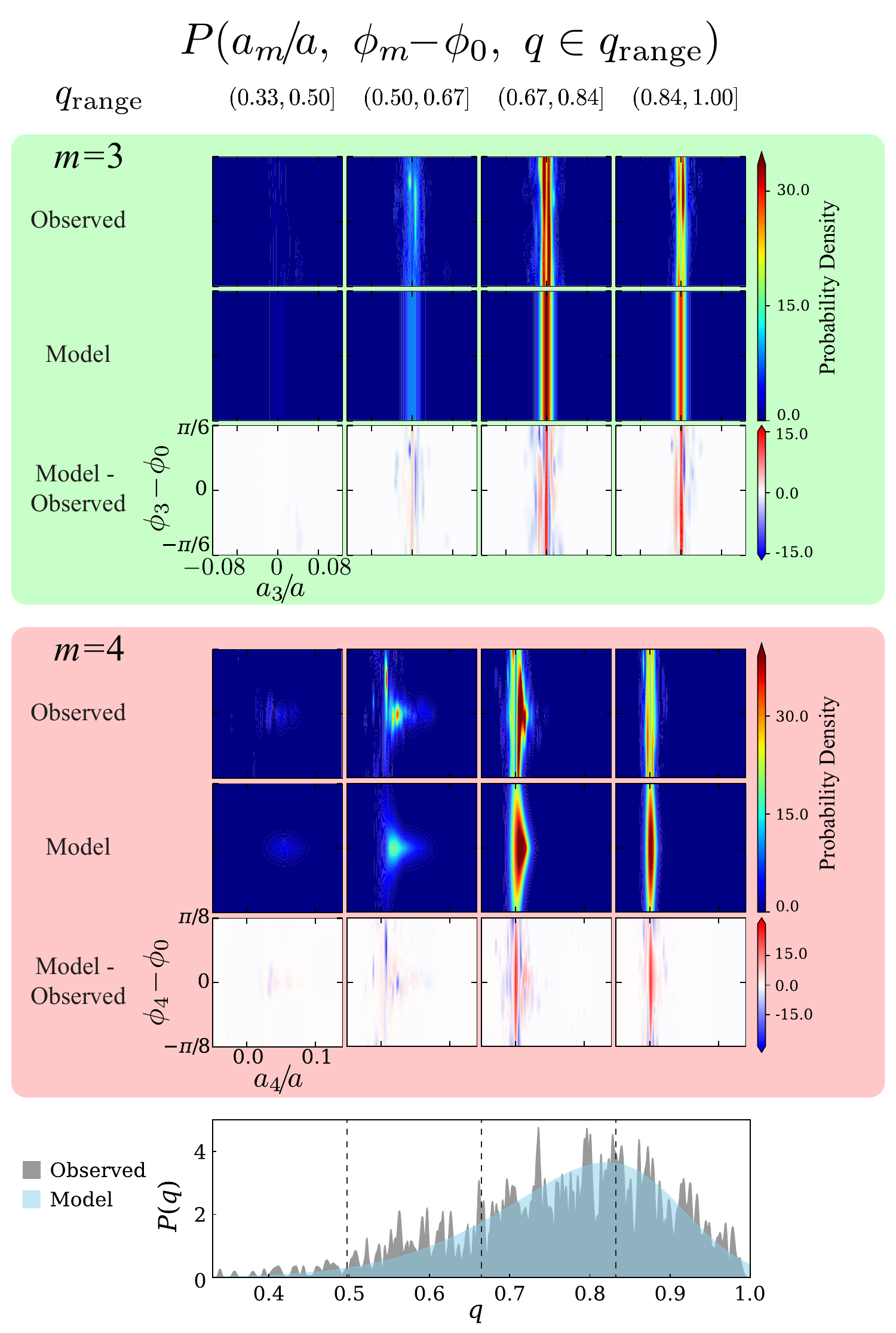}
    \caption{Observed versus modeled probability distributions for $(\X3,\Y3,\Z)$ \textbf{(top three rows)}, $(\X4,\Y4,\Z)$ \textbf{(following three rows)}, and $P(\Z)$ \textbf{(bottom)}.
    Each triplet of heatmaps shows, from top to bottom,
    the observed kernel‑density estimate (histogram of data points convolved with their uncertainties),
     the hierarchical model prediction using the best‑fit
     parameters of Table~\ref{tab:bestfit}, and their
     difference. Four columns correspond to non‑overlapping
     slices in $\Z$; they match with the four ranges shown in $P(\Z)$ with dashed line. Axis ticks are shared between heat maps and thus omitted from the panels to reduce visual clutter except one. Coordinate axes are consistent within each set of panels for $m = 3$ and $m = 4$.}
    \label{fig:prior_validation}
\end{figure}

H06 has shown that $m=3$ and $m=4$ multipoles have different distributions\footnote{More exactly, H06 showed that $\alpha_4\!/\!a$ tends to have asymmetric distribution toward the positive, whereas $\beta_4\!/\!a$, $\alpha_3\!/\!a$, and $\beta_3\!/\!a$ do not. See Appendix \ref{appendix:multipole_convention} for the comparison of two multipole conventions, $(\X{m}, \Y{m})$ and $(\alpha_m\!/\!a, \beta_m\!/\!a)$.}. However, plotting $\X{m}$ together with $\Y{m}$ with different $\Z$ ranges reveals more correlations from the data of H06.
See ``Observation'' columns of Figure~\ref{fig:prior_validation} for the distribution of $(\X3, \ \Y3)$ and $(\X4, \ \Y4)$ with different $\Z$ ranges from the isophotes of 840 E/S0 galaxies\footnote{The original number of data points is 847, but 7 of them were disposed in conversion and error analysis; see Appendix~\ref{app:obs_kde}}, with the fourth power of velocity dispersion ${v_\text{dis}}^4$ as weight\footnote{$\theta_E\propto {v_\text{dis}}^2$ for a singular isothermal sphere, and thus the strong lensing area $\pi {\theta_E}^2 \propto {v_\text{dis}}^4$; refer to \cite{narayan1996lectures}}, of which the original data is provided by H06. Four important correlations can be observed:
\begin{enumerate}
    \item $\X4$ tends to be positive and greater (more disky) when $\Z$ is smaller (more elliptical).
    \item $\Y4$ tends to be close to $0$ (more aligned) when $\X4$ is positive \& greater (more disky).
    \item $\X3$ is symmetric around $0$ regardless of $\Z$; whereas the width of the true distribution varies.
    \item $\Y3$ does not have a strong correlation with $\X3$ or $\Z$.
\end{enumerate}

Based on the behavior of the observed population density, we model the joint probability density of the multipole variables as
\begin{equation}
    \label{eq:factorization1}
    \begin{split}
            P(\X{m},\ \Y{m},\ \Z) 
    &= P(\Z) \; P(\X{m},\ \Y{m} \ |\ \Z)
    \\
    &= P(\Z) \; P(\X{m}\ |\ \Z) \; P(\Y{m} \ |\ \X{m})
    \end{split}
\end{equation}
by assuming that \Y{m} is conditionally independent of \Z, given \X{m}: 
\(
P(\Y{m}\ |\ \X{m}, \ \Z) = P(\Y{m}\ |\ \X{m})
\). There is no meaningful correlation between the two multipole pairs $(\X3,\Y3)$ and $(\X4,\Y4)$, so we assume conditional independence between them to build the probability density of the five variables:
\begin{equation}
    \label{eq:factorization2}
    \begin{split}
    P(\X3,\ \Y3,\ \X4,\ \Y4,\ \Z) 
    = P(\Z) \; & P(\X3,\ \Y3 \ |\ \Z) \; P(\X4,\ \Y4 \ |\  \Z)
    \\
    = P(\Z)\; & P(\X3 \ |\ \Z) \;P(\Y3\ | \ \X3) \\
    & P(\X4 \ |\ \Z) \;P(\Y4\ | \ \X3)
\end{split}
\end{equation}

The functional choices are:

\begin{itemize}
  \item $P(\Z)$ — skew–normal distribution with three parameters:
        shape $\alpha_\Z$, location $\xi_\Z$, and scale $\omega_\Z$. Normalized within $\left(\Z_\text{min}=0.33,\ \Z_\text{max}=1\right)$ such that $\int_{\Z_\text{min}}^{\Z_\text{max}} P(\Z) \ \d \Z = 1$.
  \item $P(\X3 \mid \Z)$ — Gaussian distribution with fixed mean $\mu = 0$ and standard deviation $\sigma(\Z)$,
    where $\sigma(\Z)$ is defined by linear interpolation through two control points. Normalized within $\left( (\X3)_\text{min}=-0.08, \ (\X3)_\text{max}=0.08 \right)$.
  \item $P(\Y{3} \mid \X3)$ — uniform distribution over the interval $[(\Y{3})_\text{min}=-\pi/6,\,(\Y{3})_\text{max}=\pi/6)$.
  \item $P(\X4 \mid \Z)$ — skew–normal distribution with shape $\alpha(\Z)$, location $\xi(\Z)$, and scale $\omega(\Z)$,
        each specified by a three-point linear spline. Normalized within $\left( (\X4)_\text{min}=-0.05, \ (\X4)_\text{max}=0.14 \right)$.
  \item $P(\Y4 \mid \X4)$ — symmetric generalized Gaussian distribution with fixed mean $\mu = 0$, and
        scale $\alpha(\X4)$ and shape $\beta(\X4)$, both described by three-point linear splines. Normalized within $[(\Y{4})_\text{min}=-\pi/8, \,(\Y{4})_\text{max}=\pi/8)$.
\end{itemize}

The limits
$\Z_\text{max}=1$, $(\Y{m})_\text{min}=-\pi/2m$, and $(\Y{m})_\text{max}=\pi/2m$ are set by their mathematical definition, whereas $\Z_\text{min}$, $(\X{m})_\text{min}$, $(\X{m})_\text{max}$,
are set by the observed population limit; more specifically, maximum/minimum of data value $\pm$ 3 times of uncertainty.

The parameters were optimized by minimizing the sum of two Jensen–Shannon (JS) divergences between the model and the galaxy sample from \cite{hao2006isophotal} convolved with its uncertainties. The final JS divergence for $m=3$ and $m=4$ are $\mathrm{JS}_{m=3}=0.100$ and $\mathrm{JS}_{m=4}=0.143$, respectively.
The details of optimization is shown in Appendix~\ref{app:optimization}. Table~\ref{tab:bestfit} lists the best–fit values that we use throughout this work.

\begin{table}[b]
\centering
\caption{Best–fit hyperparameters of the hierarchical prior. The total number of free parameters to be optimized are 37.}
\label{tab:bestfit}
\begin{tabularx}{\textwidth}{@{}l l X@{}}
\toprule
\textbf{Group} & \textbf{Symbol} & \textbf{Value(s)} \\
\midrule
\multirow{3}{*}{$P(\Z)$}          & $\alpha_\Z$ & $-2.491$ \\
                                  & $\xi_\Z$    & $0.910$ \\
                                  & $\omega_\Z$ & $0.175$ \\
\midrule
\multirow{2}{*}{$\sigma(\Z)$ spline ($m{=}3$)} &
$\sigma_{x,3}$ & $(0.482,\;0.871)$ \\
                                  & $\sigma_{y,3}$ & $(0.0075,\;0.0043)$ \\
\midrule
\multirow{2}{*}{$\alpha(\Z)$ spline ($m{=}4$)} &
$\alpha_{x,4}$ & $(0.522,\;0.614,\;0.838)$ \\
                                  & $\alpha_{y,4}$ & $(0.872,\;-1.054,\;-0.224)$ \\
\midrule
\multirow{2}{*}{$\xi(\Z)$ spline ($m{=}4$)} &
$\xi_{x,4}$    & $(0.437,\;0.678,\;0.851)$ \\
                                  & $\xi_{y,4}$    & $(0.0485,\;0.0124,\;0.0008)$ \\
\midrule
\multirow{2}{*}{$\omega(\Z)$ spline ($m{=}4$)} &
$\omega_{x,4}$ & $(0.533,\;0.628,\;0.942)$ \\
                                  & $\omega_{y,4}$ & $(0.0195,\;0.0110,\;0.0038)$ \\
\midrule
\multirow{2}{*}{$\alpha(\X4)$ spline} &
$\alpha_{x,4}'$ & $(0.0037,\,0.0123,\,0.0340)$ \\
                     & $\alpha_{y,4}'$ & $(0.540,\;0.139,\;0.068)$ \\
\midrule
\multirow{2}{*}{$\beta(\X4)$ spline} &
$\beta_{x,4}$ & $(0.0045,\;0.0131,\;0.0229)$ \\
                                  & $\beta_{y,4}$ & $(1.929,\;1.175,\;1.601)$ \\
\bottomrule
\end{tabularx}
\end{table}

Figure~\ref{fig:prior_validation} compares the observed distributions
in the \cite{hao2006isophotal} catalog (left panels), the best‑fit model
predictions (center), and their residuals (right) for both
$m=3$ and $m=4$ multipoles in four axis‑ratio slices. The general agreement between the observed and best-fit model prediction demonstrates that the
chosen functional forms are flexible enough to capture the empirical
population while retaining analytic simplicity.

\begin{figure}[H]
    \centering
    \includegraphics[width=\linewidth]{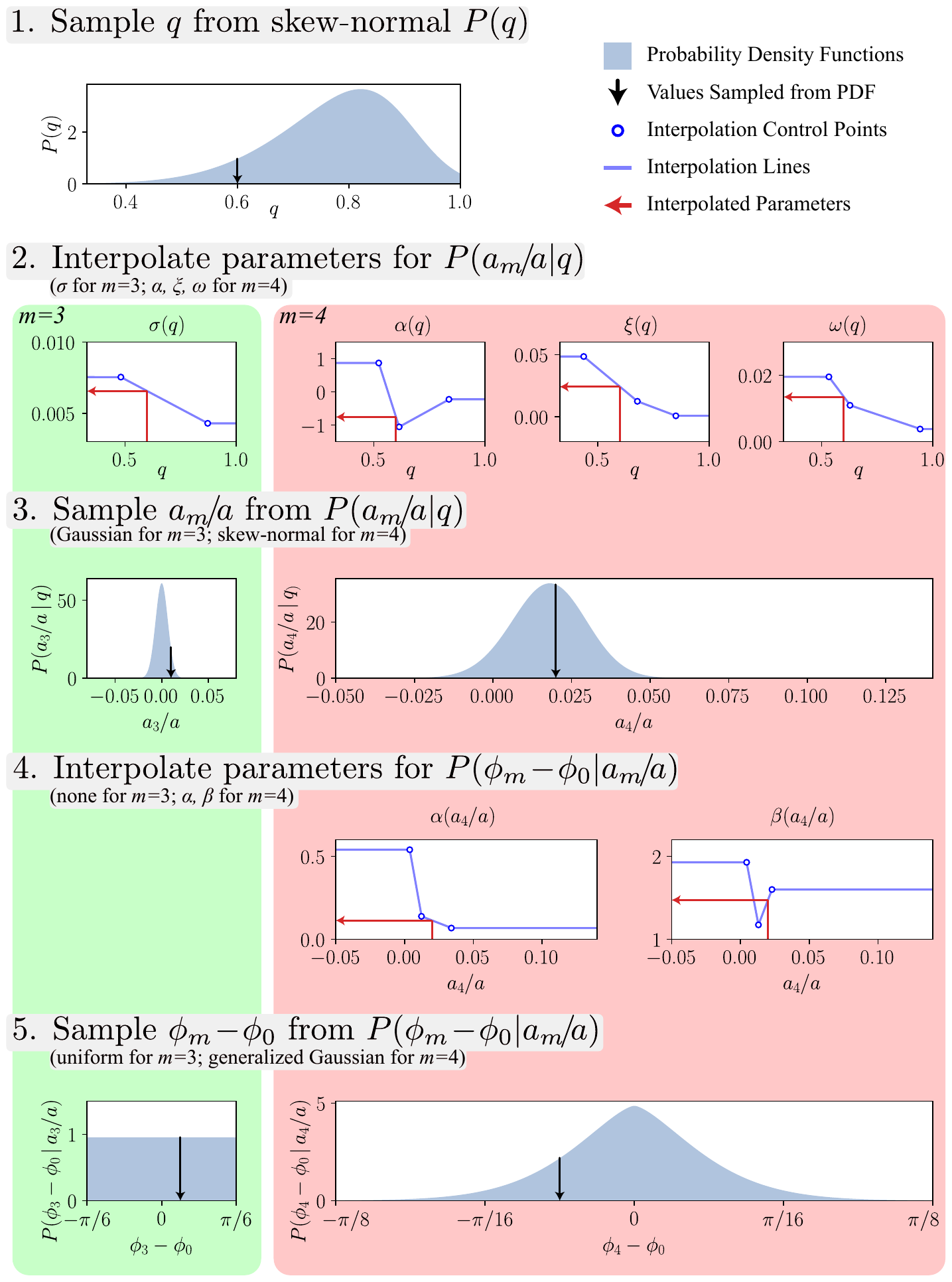}
    \caption{Graphical illustration of the hierarchical prior and
             the sampling steps summarized in
             Algorithm~\ref{alg:sampling}.  
             Blue shaded curves show probability density functions; black arrows denote the sampled values; blue dots and lines are interpolation control points and interpolated values, respectively; red lines represent parameter interpolation.}
    \label{fig:prior_design}
\end{figure}

Algorithm~\ref{alg:sampling} and Figure~\ref{fig:prior_design} demonstrate the sampling procedure. In our inference pipeline we do not draw these parameters directly; instead the joint prior of Eq.~\eqref{eq:factorization2} is encoded as an additive log‑likelihood evaluated for each MCMC proposal.

\noindent
\begin{minipage}{\linewidth}
\begin{algorithm}[H]
\setlength{\columnsep}{0.5em}
\setlength{\columnseprule}{0pt}
\caption{Draw a sample $(\X3,\,\Y3,\,\X4,\,\Y4,\,\Z)$. The same factorization is implemented as a custom prior‑likelihood term to be fed into \textsc{lenstronomy} and \textsc{emcee} pipeline. The sampling steps below show the conceptual sampling equivalent.}
\label{alg:sampling}
\small
\begin{enumerate}[label=\arabic*., leftmargin=1.2em]
\item Sample $\Z \sim \mathrm{SkewNormal}(\alpha_\Z,\xi_\Z,\omega_\Z)$.
\item
  \begin{enumerate}[label=\alph*.]
  \item Compute $\sigma(\Z)$ by linear interpolation of the two
        control points $(\sigma_{x,3},\sigma_{y,3})$.
  \item Sample $\X3 \sim \mathcal{N}(0,\sigma(\Z)^2)$.
  \item Sample $\Y3 \sim \mathcal{U}(-\pi/6,\pi/6)$.
  \end{enumerate}
\item
  \begin{enumerate}[label=\alph*.]
  \item Interpolate $(\alpha,\xi,\omega)(\Z)$ from their control
        points to obtain the shape of $P(\X4|\Z)$.
  \item Sample $\X4 \sim \mathrm{SkewNormal}(\alpha(\Z),\xi(\Z),\omega(\Z))$.
  \item Interpolate $\alpha(\X4)$ and $\beta(\X4)$ from their control
        points.
  \item Sample $\Y4$ from the symmetric generalized Gaussian
        $G(\mu\!=\!0, \alpha(\X4),\beta(\X4))$ on $[-\pi/8,\pi/8)$.
  \end{enumerate}
\end{enumerate}
\end{algorithm}
\end{minipage}

\vspace{1em}

\newpage

\newpage

\section{Mock–data construction}
\label{sec:mock}

\begin{table}[b]
    \centering
    \begin{tabular}{c c c l } 
    \hline 
          \makecell{\hspace{0.1cm}Multipole Scenario\hspace{0.1cm}}  &  \makecell{\hspace{0.1cm}Parameter Name\hspace{0.1cm}}&  \makecell{\hspace{0.1cm}Simulation Truth\hspace{0.1cm}} & \makecell{\hspace{0.1cm}Fitting Prior\hspace{0.1cm}} \\ \Xhline{3\arrayrulewidth}
         \multirow{4}{*}{\makecell{Mild $m=4$}} 
         & $\X{3}$ & $0.01$  &   \multirow{4}{*}{\makecell{Jointly sampled\\(see Section~\ref{sec:prior})}} \\
         & $\Y{3}$ & $ 0.2 $ (rad)  &  \\
         & $\X{4}$ & $0.01$ & \\ 
         & $\Y{4}$ & $ 0 $ (rad) &  \\ 
         \hline 
         \multirow{4}{*}{\makecell{Strong $m=4$}} 
         & $\X{3}$ & $0.01$  &   \multirow{4}{*}{\makecell{Jointly sampled\\(see Section~\ref{sec:prior})}} \\
         & $\Y{3}$ & $ 0.2 $ (rad) &  \\
         & $\X{4}$ & $0.03$ &  \\ 
         & $\Y{4}$ & $ 0 $ (rad) &  \\
         \hline 
    \end{tabular}
    \caption{$m\!=\!3,4$ multipole parameters with two different scenarios. 
    }
    \label{table:m=34_true_params}
\end{table}

Our goal is to provide a controlled, HST–quality
reference case in which the impact of multipole priors and
extended–arc information can be quantified unambiguously. Therefore, we fix the macro model parameters and source morphology based on a single,
well-studied system with clear extended arcs: WGD\,2038-4008
\cite{agnello_2018,shajib2019every}.
This system has a source redshift of $z_{\text{source}}=0.777$, $z_\text{lens}=0.230$ \citep{agnello_2018,krone2018gaia}. We fitted the original HST F814W image of this system with a model without multipoles---EPL+shear for lens mass, Sérsic for lens light and source light profiles, and point-like quasars---to set the realistic model parameters. This ensures that in addition to realistic signal to noise, our mock lenses will have typical properties of gravitational lenses including potential contamination from the deflector light and realistic extended quasar host galaxy brightness. We add realistic mass multipole parameters on the fitted model based on two multipole scenarios. Investigation for multiple mock/real lens systems is left to future work. Below we summarize the ingredients.

\subsection{Parameter setting}
\label{subsec:mock_params}

\paragraph{Multipole scenarios}
We base our parameter choices within a reasonable prior shown in Section \ref{sec:prior}.
We test two scenarios for the multipoles of the main deflector as described in Table \ref{table:m=34_true_params}. 
One scenario, `Mild $m=4$' has $\X{4}=0.01$. The other scenario, `Strong $m=4$' has $\X{4}=0.03$. In both cases, we align the $m=4$ multipole with EPL; $\Y4=0\text{ (rad)}$, while keeping the prior broad. For $m=3$ multipoles, we set them to be $\X{3}=0.01$ and $\Y{3}=0.2\text{ (rad)}$ for both cases.
\par
These multipole scenarios are selected as they are representative of realistic combinations of parameter values seen in galaxy isophotes, and they span a possible range of perturbations to the lens models.

\begin{figure}[b]
\centering\includegraphics[scale=0.6]{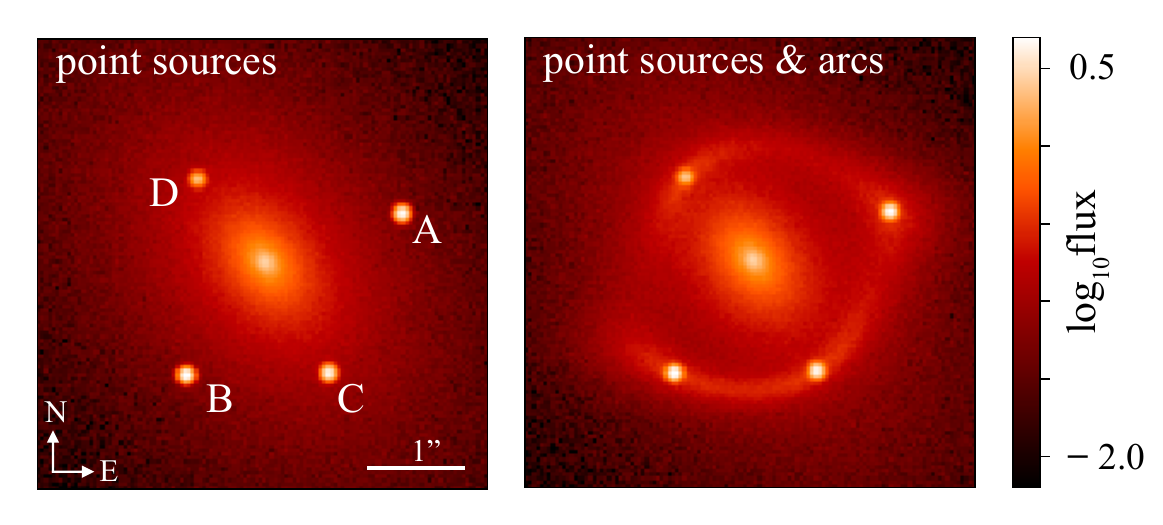}
    \caption{Comparison of simulations of HST-quality quadruply imaged quasars with point sources only \textbf{(left)} and point sources and extended arcs \textbf{(right)}. The four quasar images are marked as A, B, C, and D. Note that the `point sources only' case uses limited information (point source positions) only, whereas the `point sources and extended arcs' case corresponds to an analysis using complete information with extended arcs. This way we can measure how much improvement comes when the full observation with extended arcs is modeled, compared to using images positions only.
    The change from one multipole scenario to another is quite subtle (e.g. slight shift of point sources and arcs), and thus not compared here.
    } 
    \label{fig:ps_vs_ps+arc}
\end{figure}

\paragraph{Observation scenarios}
For each multipole scenario described above, we made two simulated observations of the lens system; one in which only point sources were imaged, and the other with point sources and extended arcs. The mock observation data was generated using \textsc{lenstronomy}.
Figure \ref{fig:ps_vs_ps+arc} gives an example of two of our mock data sets.
\par

\paragraph{Other parameters}
The parameter values and priors for other mass and light components are summarized in Tables~\ref{table:lens_mass_params} and \ref{table:light_params}.
Note that, for the \emph{point–sources only} data sets the host light is not included.

\subsection{Data quality}
\label{subsec:data_quality}
We create mock observations based on Hubble Space Telescope's WFC3 imaging data from programs GO-15320 and GO-15652 (PI: Treu, T.). Those programs carried out a uniform multi-band imaging campaign of 31 quadruply imaged quasars and \cite{shajib2019every} provided the optimal parameter for modeling our mock lens system, WGD 2038-4008. We base our simulations on the observations with F814W which had the best combination of sensitivity and clear extended arcs.
\par
The exposure time is set to be 920 seconds, the pixel size is set to be $0.04$ arcseconds, and the background noise level is set to be $0.006$ photons/second for each pixel. 
The point spread function (PSF) is modeled as a two-dimensional Gaussian with a known full width at half maximum (FWHM) of 0\farcs10. This choice is a simplification; in real observations, the PSF is typically more complex and must be inferred from the data. We opt not to model a complex PSF or reconstruct it from the data to avoid PSF-related complications.
Section~\ref{sec:inference} now treats these mocks as data and applies
the modeling framework of Section~\ref{sec:models}.

\section{Inference framework}
\label{sec:inference}

\subsection{Likelihood and priors}
\label{subsec:likelihood}

We infer the lens model parameters from the mock data following the standard \textsc{lenstronomy} fitting procedure \citep{birrer2018lenstronomy,birrer2021lenstronomy} together with two custom prior likelihoods. The total log-likelihood of the inference process is given as
\begin{equation}
\begin{split}
\ln\mathcal L_\mathrm{tot}
   = 
   &\  \ln\mathcal L_{\mathrm{img}}
               \;+\; \ln\mathcal L_{\mathrm{ps}}       
   \;+\;
   \underbrace{\ln \mathcal{L}_e
   }_{\text{ellipticity prior, Appendix~\ref{app:e1e2_prior}}}
   \\[5pt]
   \;&+\;
   \underbrace{\ln P_\text{model}\bigl(\X3,\Y3,\X4,\Y4,\Z\bigr)
               }_{\text{population prior, Section~\ref{sec:prior}}} 
\end{split}
\end{equation}
where:

\begin{itemize}
\item $\mathcal L_{\mathrm{img}}$ is the pixel–based image likelihood. Note that the point–source \emph{fluxes} are \emph{not} compared to the observations as done by other studies of lensed quasars using HST \citep[e.g.][]{schmidt2023strides,agnello_2018}; quasar image flux in HST optical broad-band imaging is dominated by emission from the quasar accretion disk and thus sensitive to stellar microlensing in reality.
\item $\mathcal L_{\mathrm{ps}}$ quantifies how well the multiple lensed quasar images can be mapped back to a common point in the source plane under the current lens model. 
\item $\mathcal{L}_e$ is an ellipticity prior that makes uniform prior on $(\Z, \phi_0)$.
See Appendix~\ref{app:e1e2_prior} for details.
\item $P_\text{model}$ is the joint prior for multipole parameters and $q$ described in Section~\ref{sec:prior}.
\end{itemize}
Other parameters have broad uniform priors (see Tables~\ref{table:lens_mass_params} and \ref{table:light_params}). The true PSF is assumed to be known, and thus not optimized.

\subsection{Sampling strategy}
\label{subsec:sampler}

The posterior landscape is explored in two stages:

\begin{enumerate*}[label=(\roman*)]
\item {\bf Particle–swarm optimization} (\textsc{PSO}), 100 particles,
      200 iterations, identifies a high–likelihood region.
\item {\bf Ensemble MCMC} (the {\tt emcee} sampler) with  
      a walker-to-parameter ratio of 8:1, 40\,000 steps each. The first 30\,\% of the chain is
      discarded as burn–in; out the remaining chain, ${\sim}\,500\,000$ samples are
      used for all posterior plots and flux–ratio statistics.
\end{enumerate*}

\subsection{Derived quantities}
\label{subsec:derived}

For each retained sample we evaluate the signed magnifications
\begin{equation}
    \mu(\vec{\theta}) = \frac{1}{(1-\kappa)^2 - ({\gamma_1}^2 + {\gamma_2}^2)}
    ,
\end{equation}
where
$\kappa = \frac{1}{2}\left[
\frac{\partial^2 \psi}{\partial {x}^2}
+
\frac{\partial^2 \psi}{\partial {y}^2}
\right]$,
$\gamma_1 = \frac{1}{2}\left[
\frac{\partial^2 \psi}{\partial {x}^2}
-
\frac{\partial^2 \psi}{\partial {y}^2}
\right]$,
and
$\gamma_2 =
\frac{\partial^2 \psi}{\partial {x}\partial {y}}$ for the projected gravitational potential $\psi$. The three model-predicted flux ratios were calculated as
\(
B/A = |\mu_B/\mu_A|,\;
C/A = |\mu_C/\mu_A|,\;
D/A = |\mu_D/\mu_A|.
\)
\par
For each multipole scenario, we define the with-arcs precision improvement factor, $\mathcal{F}_{(p)}$, as the ratio between the 68\% confidence interval uncertainties of a variable $p$ for `point source only' case, $\Delta p|_\text{point sources}$, and that of `point sources and arcs' case, $\Delta p|_\text{point sources \& arcs}$.
\begin{equation}
    \mathcal{F}_{p} \equiv \frac{\Delta p|_{\text{point sources}}}{\Delta p|_{\text{point sources \& arcs}}} .
\end{equation}

\section{Results}
\label{sec:results}

\begin{figure}
\centering
    \begin{tikzpicture}
    \draw (0, 0) node[inner sep=0] (FIG) {\includegraphics[width=\linewidth]{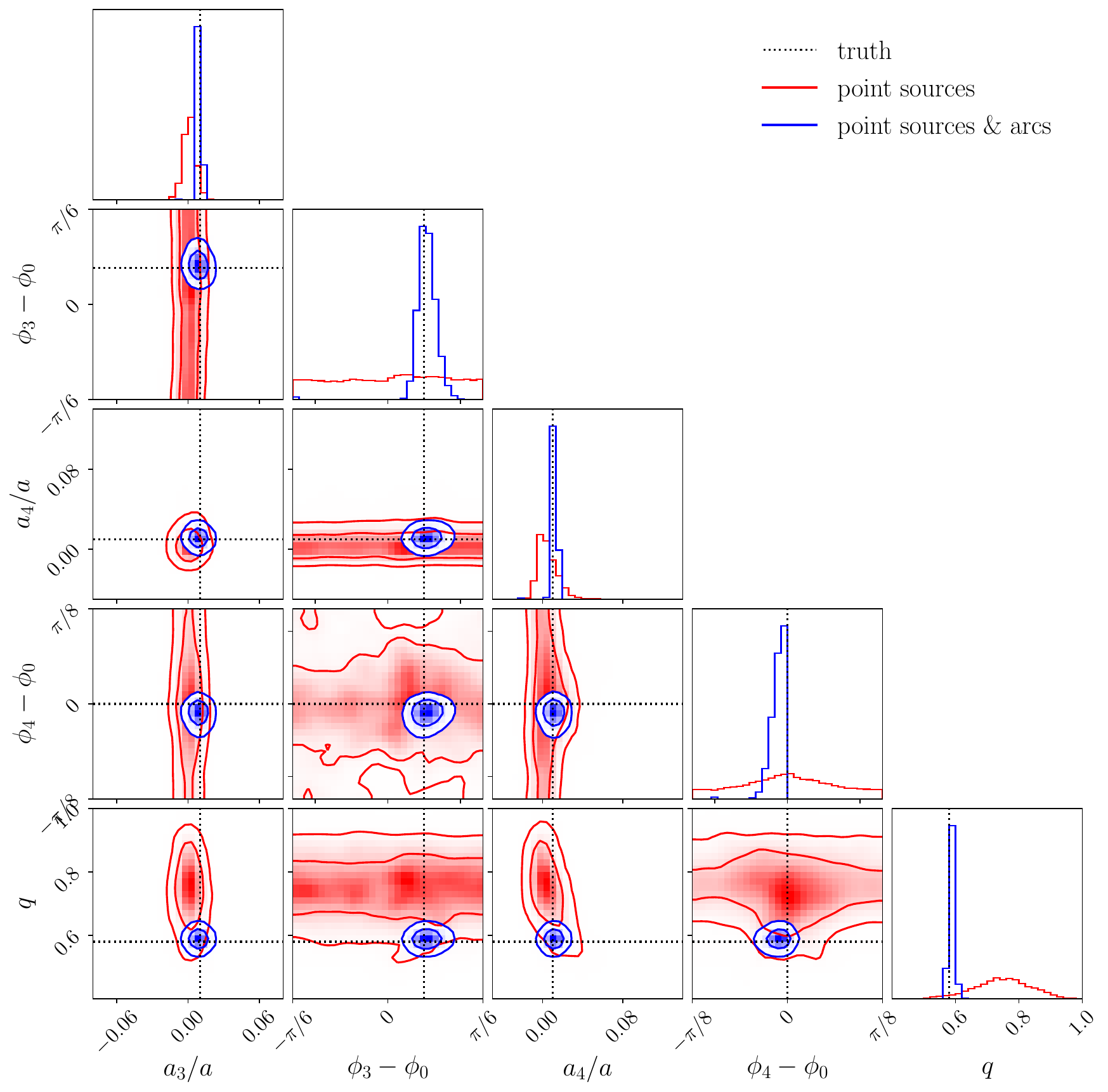}};
    \draw ([xshift=0cm,yshift=0.2cm]FIG.north) node {\large `Mild $m=4$' Inference Result};
\end{tikzpicture}
\caption{Corner plot of the joint posterior for the multipole parameters in the `Mild $m=4$' mock lens.
 Red contours/histograms correspond to the point‑source–only fit, while blue show the fit that also uses the extended arcs. Shaded regions (light to dark) enclose the 
 68\% and 95\% highest‑posterior‑density levels. Black dashed lines indicate the truth values. Numerical summaries are given in Table~\ref{tab:multipoles_inference_result}.}
 \label{fig:corner_mild_multipole}
\end{figure}

\begin{figure}
\centering
    \begin{tikzpicture}
    \draw (0, 0) node[inner sep=0] (FIG) {\includegraphics[width=\linewidth]{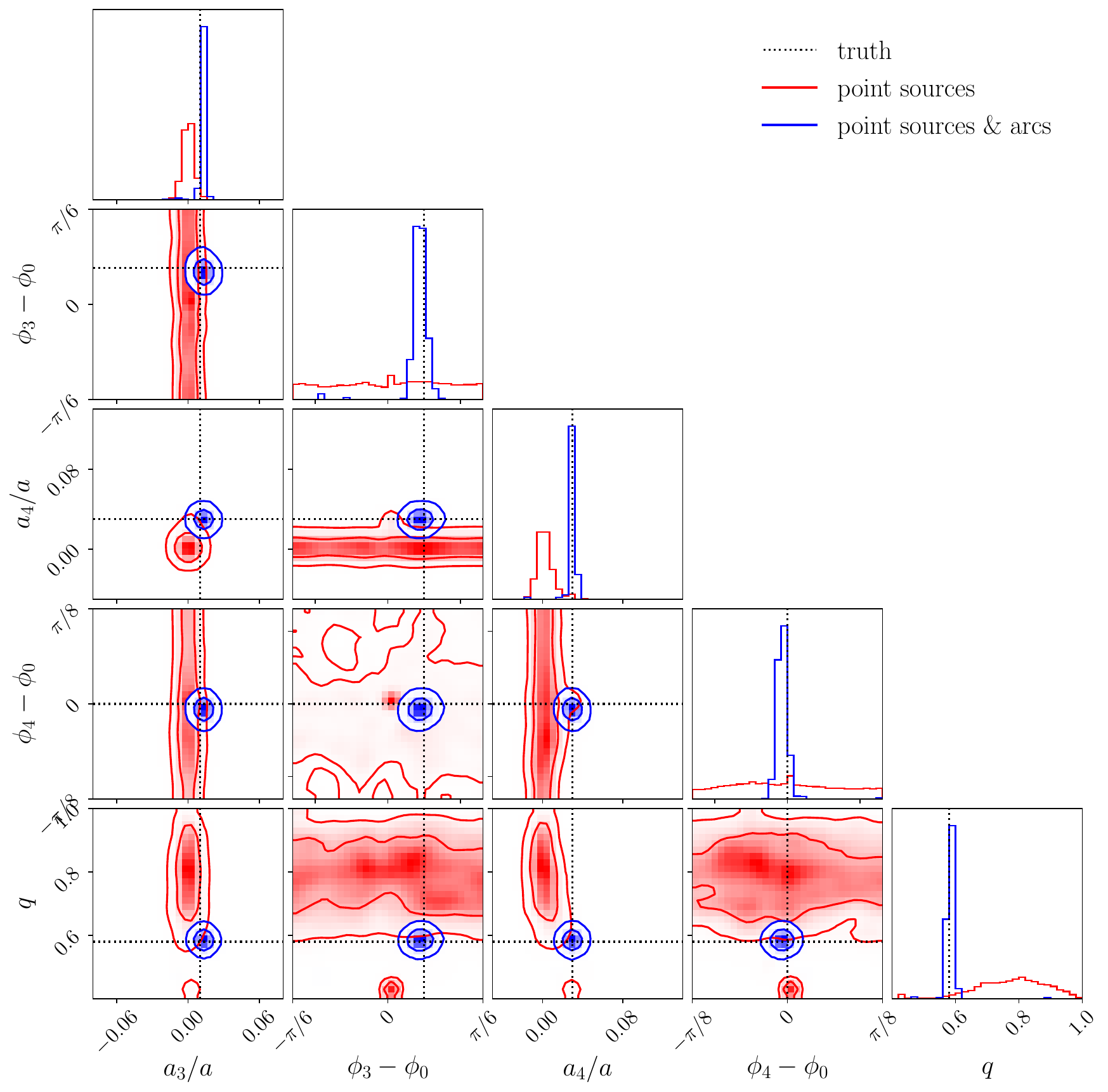}}; 
    \draw ([xshift=0cm,yshift=0.2cm]FIG.north) node {\large `Strong $m=4$' Inference Result};
\end{tikzpicture}
\caption{
Same as Figure~\ref{fig:corner_mild_multipole} but for `Strong $m=4$' mock lens. 
}
\label{fig:corner_strong_multipole}
\end{figure}

\begin{figure}
\begin{subfigure}{0.5\hsize}
\centering
    \begin{tikzpicture}
    \draw (0, 0) node[inner sep=0] (FIG) {\includegraphics[scale=0.4]{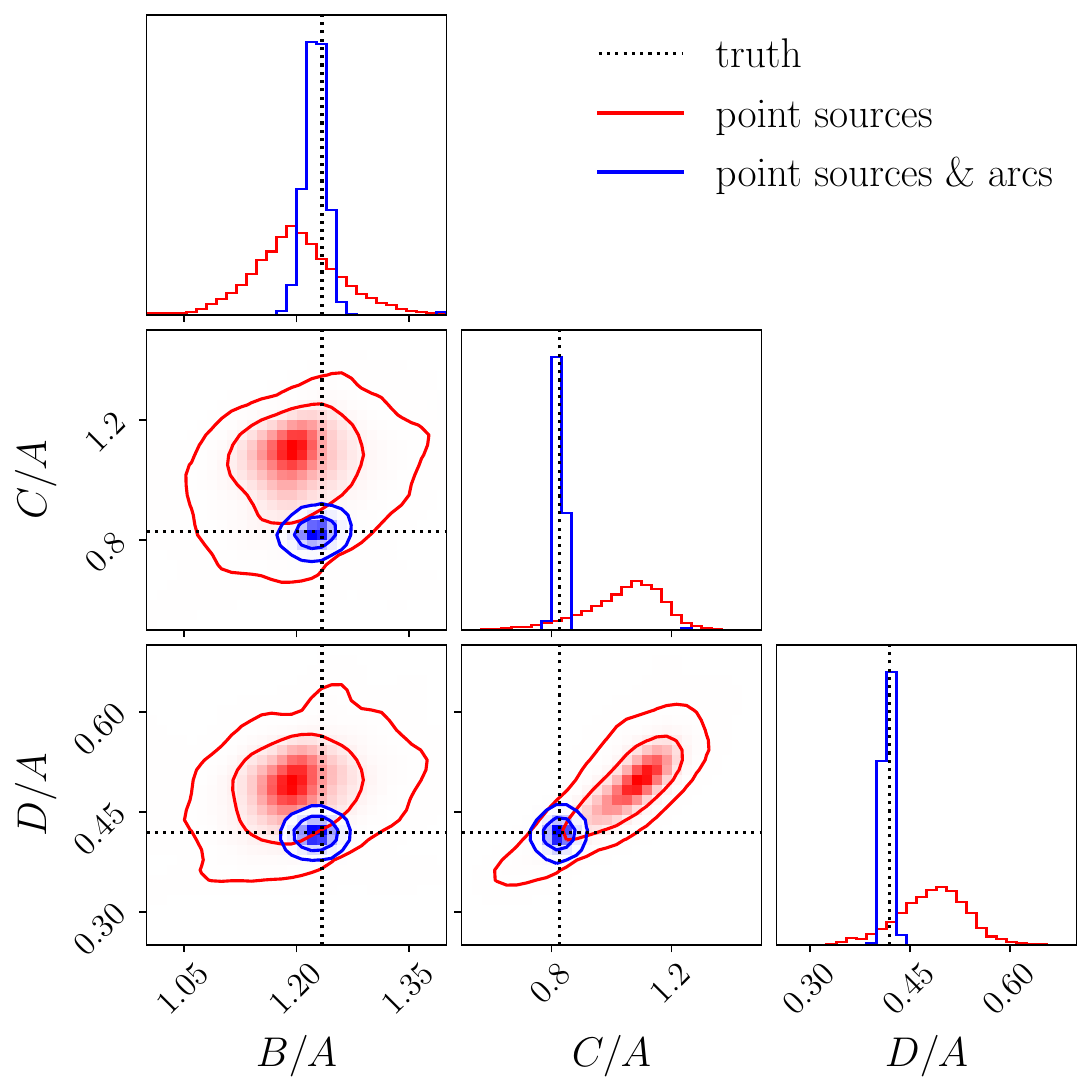}};
    \draw ([xshift=0cm,yshift=0.2cm]FIG.north) node {\large `Mild $m=4$' Inference Result};
\end{tikzpicture}
 \label{fig:corner_mild_fr}
\end{subfigure}%
\begin{subfigure}{0.5\hsize}\centering
    \begin{tikzpicture}
    \draw (0, 0) node[inner sep=0] (FIG) {\includegraphics[scale=0.4]{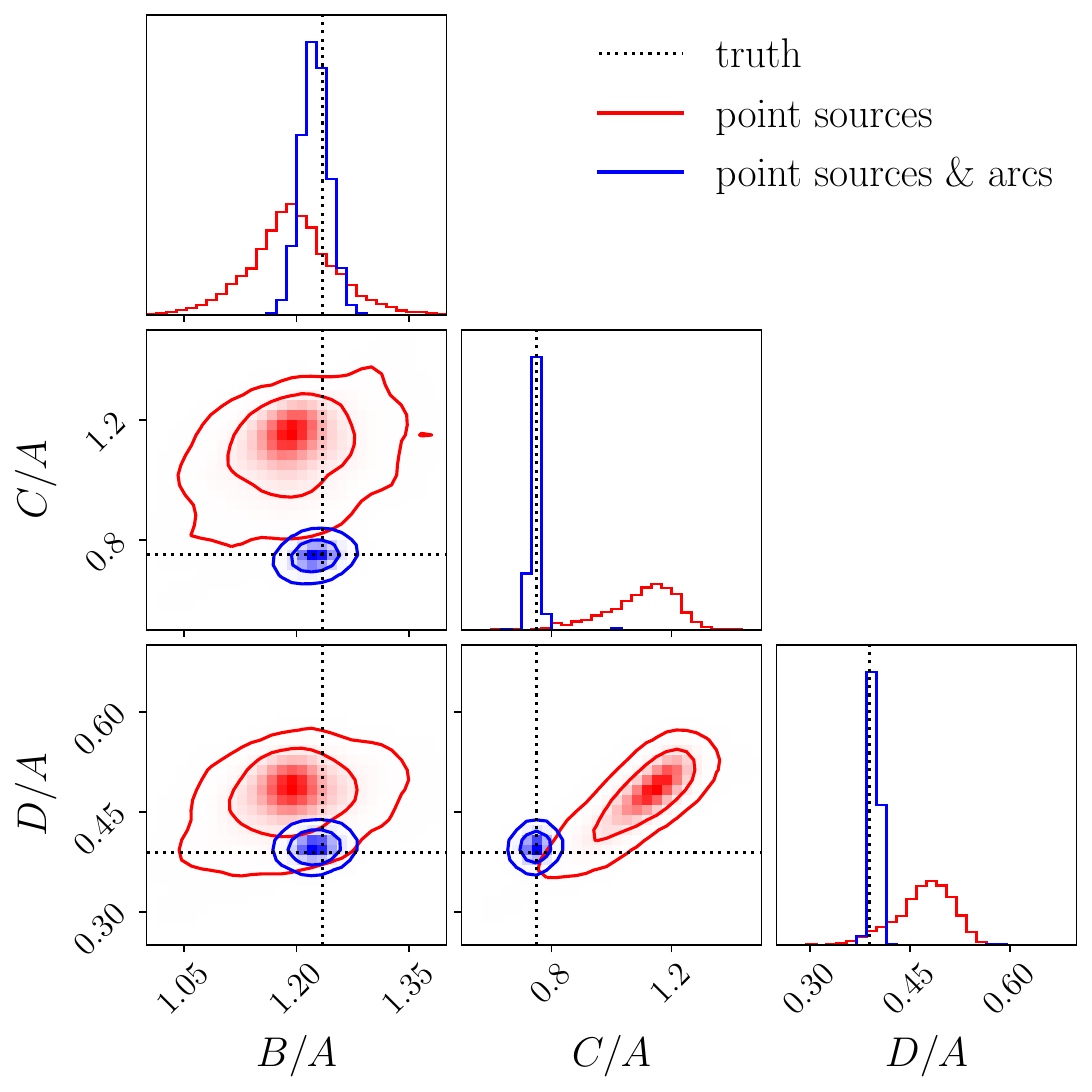}}; 
    \draw ([xshift=0cm,yshift=0.2cm]FIG.north) node {\large `Strong $m=4$' Inference Result};
\end{tikzpicture}
 \label{fig:corner_strong_fr}
\end{subfigure}
\caption{
Same as Figures~\ref{fig:corner_mild_multipole} and \ref{fig:corner_strong_multipole} but for flux ratios.
Numerical summaries are given in Table~\ref{tab:fr_inference_result}.
}
\label{fig:corner_fr}
\end{figure}

The posterior comparison of \textit{point source only} (red) and \textit{point sources + arcs} (blue) are shown in Figures~\ref{fig:corner_mild_multipole} and \ref{fig:corner_strong_multipole} for the multipole parameters and in Figure~\ref{fig:corner_fr} for the flux ratios. The improvement in parameter precision is summarized in Tables~\ref{tab:multipoles_inference_result} and \ref{tab:fr_inference_result}.

The axis ratio $\Z$ is significantly better constrained when extended arcs are included, with the precision improvement factor $\mathcal{F}_{q}$ exceeding 10 in both multipole scenarios. This more narrowly defined $\Z$ distribution in turn enables sharper joint priors on the multipole parameters.
\par
The amplitudes of the multipoles, $\X{3}$ and $\X{4}$, are largely unconstrained by point source positions alone; their posteriors remain broad and prior-dominated. When arcs are added, however, these parameters become well-constrained, with precision improvement factors of $\mathcal{F}_{\X3} \approx \mathcal{F}_{\X4} \approx 3.5$.
\par
The multipole angles $\Y{3}$ and $\Y{4}$ also show clear improvement in constraint when arcs are included. In the mild $m=4$ case, the inclusion of arcs improves the constraint on $\Y4$ by factors of
$\mathcal{F}_{\Y{3}}\approx\mathcal{F}_{\Y{4}}\approx7.5$, and in the strong $m=4$ case, the factors increase to $\approx 12$. This indicates that arc information becomes especially powerful when the underlying multipole amplitude is more pronounced.
\par
Lastly, the inclusion of arcs also greatly sharpens the predicted flux ratios. On average, the uncertainties on $B/A$, $C/A$, and $D/A$ are reduced by a factor of approximately 6, regardless of the underlying multipole scenario. This substantial improvement in predicted flux precision underscores the benefit of arc information for studies relying on flux ratio anomalies.
\par
In summary, when multipole amplitudes are at the typical level observed in the isophotes of massive elliptical galaxies, extended arcs are essential for deriving accurate and precise constraints on the smooth lens model. Point sources alone are insufficient for this task, but their combination with arcs enables strong recovery of both multipole parameters and flux ratios. This improvement comes in part from constraining the axis ratio $q$.

\begin{table}[]
    \centering
    \begin{tabular}{l c c c c c}
    \hline
    \\[-2pt]
    Multipole Parameters& $\X{3}$ & $\Y{3}$ & $\X{4}$ & $\Y{4}$ & $\Z$ 
    \\[10pt] \hline
    \multicolumn{5}{l}{Mild $m=4$}\\ \hline
    Truth & $0.010$ & $0.200$ & $0.010$ & $0.000$ & $0.580$ \\
    Point sources & $0.001^{+0.005}_{-0.005}$ & $0.036^{+0.316}_{-0.382}$ & $0.003^{+0.011}_{-0.007}$ & $0.001^{+0.214}_{-0.210}$ & $0.750^{+0.086}_{-0.090}$ \\
    Point sources \& arcs & $0.009^{+0.001}_{-0.001}$ & $0.214^{+0.050}_{-0.042}$ & $0.012^{+0.002}_{-0.002}$ & $-0.034^{+0.023}_{-0.033}$ & $0.588^{+0.008}_{-0.007}$ \\
    \makecell[l]{With-arcs precision\\improvement factor} & $3.6$ & $7.5$ & $4.6$ & $7.6$ & $11.6$ \\ \hline
    \multicolumn{5}{l}{Strong $m=4$}\\ \hline
    Truth & $0.010$ & $0.200$ & $0.030$ & $0.000$ & $0.580$ \\
    Point sources & $0.000^{+0.005}_{-0.005}$ & $0.027^{+0.322}_{-0.372}$ & $0.002^{+0.008}_{-0.006}$ & $-0.017^{+0.254}_{-0.228}$ & $0.786^{+0.101}_{-0.111}$ \\
    Point sources \& arcs & $0.013^{+0.001}_{-0.001}$ & $0.176^{+0.033}_{-0.029}$ & $0.030^{+0.002}_{-0.002}$ & $-0.022^{+0.020}_{-0.019}$ & $0.584^{+0.009}_{-0.008}$ \\
    \makecell[l]{With-arcs precision\\improvement factor} & $3.1$ & $11.2$ & $3.5$ & $12.6$ & $12.2$ \\ 
    \hline
    \end{tabular}
    \caption{Comparison of truth values, inferred parameters, and precision improvements of multipole parameters for the ``Mild $m=4$'' and ``Strong $m=4$'' scenarios.
    Each row shows the true value, the inferred median and 68\% confidence intervals for the point sources only and point sources \& arcs cases, and the resulting precision improvement factor from adding arcs. This factor is defined as the ratio of the confidence interval width in the point-source-only case to that in the combined case.}
    \label{tab:multipoles_inference_result}
\end{table}

\begin{table}[]
    \centering
    \begin{tabular}{l c c c}
    \hline
    \\[-2pt]
    Flux Ratios & $B/A$ & $C/A$ & $D/A$ 
    \\[10pt] \hline
    \multicolumn{3}{l}{Mild $m=4$}\\ \hline
    Truth & $1.234$ & $0.828$ & $0.419$ \\
    Point sources & $1.199^{+0.060}_{-0.054}$ & $1.061^{+0.100}_{-0.150}$ & $0.488^{+0.046}_{-0.053}$ \\
    Point sources \& arcs & $1.226^{+0.014}_{-0.014}$ & $0.826^{+0.013}_{-0.013}$ & $0.417^{+0.007}_{-0.006}$ \\
    \multirow{2}{*}{\makecell[l]{With-arcs precision\\improvement factor}} & $4.1$ & $9.5$ & $7.7$ \\ 
    & \multicolumn{3}{c}{\cellcolor{LighterGray}
    Average improvement factor: $6.7$ 
    }
    \\[3pt]
    \hline
    \multicolumn{3}{l}{Strong $m=4$}\\ 
    \hline
    Truth & $1.235$ & $0.751$ & $0.389$ \\
    Point sources & $1.192^{+0.055}_{-0.057}$ & $1.118^{+0.090}_{-0.138}$ & $0.478^{+0.037}_{-0.048}$ \\
    Point sources \& arcs & $1.225^{+0.019}_{-0.018}$ & $0.745^{+0.013}_{-0.012}$ & $0.397^{+0.006}_{-0.006}$ \\
    \multirow{2}{*}{\makecell[l]{With-arcs precision\\improvement factor}} & $2.997$ & $9.045$ & $6.817$ \\ 
    & \multicolumn{3}{c}{\cellcolor{LighterGray}
    Average improvement factor: $5.7$ 
    }
    \\[3pt] \hline
    \end{tabular}
    \caption{Same as Table~\ref{tab:multipoles_inference_result} but of flux ratios. The average improvement factor (gray cells) summarizes the relative gain in precision for the flux ratio predictions.
    }
    \label{tab:fr_inference_result}
\end{table}

\section{Discussion and conclusion}
\label{sec:discussion}

Extended arcs are essential to constrain multipole parameters. Adding the host–galaxy arcs improves the precision of the model flux ratios by an order of magnitude and reveals the multipole amplitude and orientation.

This work provides a reasonable prior on multipole parameters for the macro model, which can increase the robustness of strong lens modeling by incorporating well-motivated complexity. Future strong lens analyses---including time-delay measurements and substructure studies---can build on this framework. In particular, for studies that use flux ratio anomalies to infer substructures \cite{Gilman2018,Gilman2019-1,Gilman2020,hsueh_2019}, our results suggest that incorporating extended arcs can improve both the precision and robustness of the analysis, while giving meaningful constraints on multipole parameters.

We adopted the circular multipole convention to remain consistent with H06, but we regard the elliptical convention as the preferred choice for future work because it better reflects realistic galaxy shapes as discussed in Section~\ref{subsec:multipoles} \cite{Ciambur_2015,paugnat2025ellipticalmultipoles}. Circular and elliptical multipoles coincide when $q=1$, and deviate only mildly when either $q\!\sim\!1$ or the multipole amplitudes are small ($|\X{m}|\!\ll\!1$). In our sample, the majority of systems satisfy $|\X{m}|\!<\!0.02$ (98.5\% for $m=3$, 87.6\% for $m=4$, 86.7\% together), while noticeably high values of $\X{4}$ appear in highly elliptical lenses. Consequently, the circular multipole prior presented here may serve as a temporary substitute for an elliptical multipole prior, particularly systems with $q\!\sim\!1$, until future optical surveys measured in the elliptical multipole convention enable the construction of a dedicated prior.

An additional implication is about the issue of shear bias. It has been shown that the external shear fitted to strong lensing systems often compensates for missing \( m=4 \) multipole terms rather than representing true cosmic shear \cite{etherington2023strong}\footnote{The truncation of the lens mass map has also been noted as a factor influencing shear measurements; see \cite{van2020impact}.}. Including \( m=4 \) multipoles in the lens model has been proposed as one solution to the lack of correlation between external shear measured in strong lens systems and cosmic shear measured in weak lensing. Our findings imply that multipole terms can be accurately constrained when arcs and a joint prior on multipoles are included. This approach could help eliminate the shear bias in future studies.

There are, however, limitations and opportunities for further development. This proof-of-concept study is based on a single lens geometry and assumes a Gaussian PSF. Several future directions are clear: (i) applying the method to multiple mock lenses with different image configurations and source morphologies, (ii) applying it to quadruply lensed quasars with arcs observed with HST, (iii) incorporating a more complex source light model, and (iv) combining the method with JWST mid-IR flux ratios in a joint macro + substructure analysis. It is important to note that our current analysis has several simplifications and does not include dark matter substructures in the mock observations or in the modeling. Therefore, although we observe significant improvement in constraining the lens model using imaging data and arcs, this does not guarantee the same level of improvement when dark matter substructures are present.

A companion paper \cite{Gilman++24} extends this work by incorporating full realizations of dark matter substructures and modeling multipoles of 
$m=3,4$ within a comprehensive lensing framework, while the joint multipole prior proposed in this study was not used. Their inference is performed directly on imaging arcs using shapelet-based source reconstructions with substructure lensing together. However, the use of highly flexible source models and inclusion of dark matter substructure introduces significant degeneracies between lens and source parameters. As a result, the posterior constraints on the multipole parameters are noticeably broader. This suggests that flexible source and inclusion of dark matter substructures can dilute the constraining power of arcs on the lens potential unless additional priors or constraints are imposed. In contrast, our controlled mock setup---with a simple source and clean arc morphology---highlights the maximum potential gain from arcs in constraining multipoles under idealized conditions.

In summary, in this work, the population prior for \((\X3,\Y3,\X4,\Y4,\Z)\) based on optical observations is designed and implemented into the strong lens system. When combined with extended arcs, yields an order-of-magnitude improvement in the precision of smooth-model multipole parameters and flux ratios. This approach may also contribute to future strong lensing studies and enable separation of line-of-sight (cosmic) external shear from apparent “shear” that is actually produced by higher-order multipoles in the deflector.

\acknowledgments
This research was conducted using MERCED cluster (NSF-MRI, \#1429783) and Pinnacles (NSF MRI, \#2019144)
at the Cyberinfrastructure and Research Technologies (CIRT) at University of California, Merced. 
We thank Caina Hao and Shude Mao for providing the original data from their optical survey of galaxy multipoles, Anowar Shajib for providing model parameters for WGD2038-4008, and Lyne Van de Vyvere for a discussion about multipole notation conventions. AMN and MO acknowledge support by the NSF through grant 2206315 "Collaborative Research: Measuring the physical properties of dark matter with strong gravitational lensing", and from \#GO-2046 which was provided by NASA through a grant from the Space Telescope Science Institute, which is operated by the Association of Universities for Research in Astronomy, Inc., under NASA contract NAS 5-03127.
DG acknowledges support from a Brinson Prize Fellowship. SB acknowledges support from the Department of Physics \& Astronomy, Stony Brook University.

\newpage
\appendix

\section{Comparison of multipole conventions}
\label{appendix:multipole_convention}

In this paper, the multipole radial deviation of the isophotal or isodensity contour from the best-fit ellipse was expressed as a single cosine function with a multipole phase $\phi_m$ as follows.

\begin{samepage}
\begin{equation}
\begin{split}
    \delta r &= a_m \cos \left(m(\phi - \phi_m)\right)
\end{split}
\end{equation}
\noindent\makebox[\linewidth]{\hfill \(\displaystyle\cdots(a_m, \phi_m) \ \mathrm{convention}\)}
\end{samepage}
\\[5pt]
Here, $r$ and $\phi$ are the typical radial and angular coordinates, $r=\sqrt{x^2 + y^2}$ and $\phi=\arctan\!2(y,x)$.
The same equations can be converted into a different convention following \citep{hao2006isophotal, bender1988isophote} using the sum of cosine and sine functions as follows\footnote{The equations in \cite{hao2006isophotal} do not have the angle of the ellipse $\phi_0$ because their coordinate system is aligned with the elliptical profile; i.e. $\phi_0 = 0$ by construction. Here we included it for generality of the equation. Also, they use $(a_m, b_m)$ instead of $(\alpha_m, \beta_m)$; we are using the latter to distinguish the conventions.}.

\begin{samepage}
\begin{equation}
    \displaystyle\delta r = \alpha_m \cos \left(m(\phi - \phi_0)\right) + \beta_m \sin \left(m (\phi - \phi_0)\right)
\end{equation}
\noindent\makebox[\linewidth]{\hfill \(\displaystyle\cdots(\alpha_m, \beta_m) \ \mathrm{convention}\)}
\end{samepage}
\\[5pt]
Note that when the multipole and the ellipse are aligned, $\phi_m - \phi_0 = 0$ in the first convention, $\beta_m=0$ in the second convention, and $a_m = \alpha_m$.
\par

The conversion from $(a_m, \phi_m)$ to $(\alpha_m, \beta_m)$ is given as follows, from the angle sum formula for cosine.
\begin{samepage}
\begin{equation}
    \alpha_m = a_m \cos(m(\phi_m - \phi_0)), \quad 
    \beta_m = a_m \sin(m(\phi_m - \phi_0))
\end{equation}
\noindent\makebox[\linewidth]{\hfill \(\
\cdots(a_m, \phi_m) \rightarrow (\alpha_m, \beta_m)
\)}
\end{samepage}
\\[5pt]

The other way of conversion from $(\alpha_m, \beta_m)$ to $(a_m, \phi_m)$ is not unique. We choose to do it by the following.

\begin{samepage}
\begin{equation}
    a_m = \mathrm{sign}\left(\alpha_m\right) \sqrt{{\alpha_m}^2+{\beta_m}^2}
    , 
    \quad 
    \phi_m = \phi_0 + \frac{1}{m} \arctan(\beta_m/\alpha_m)
\end{equation}
\noindent\makebox[\linewidth]{\hfill \(\
\cdots
(\alpha_m, \beta_m) \rightarrow (a_m, \phi_m)
\)}
\end{samepage}
\\[5pt]
This way of conversion lets $a_m$ keeps the sign of
\(
\alpha_m
\)
and its significance; e.g. $a_4>0$ means disky and $a_4<0$ means boxy. If a different conversion rule is used, this property is not guaranteed. For example, assume the following conversion: $a_m'=\sqrt{{\alpha_m}^2+{{\beta_m}^2}}$ and $\phi_m' = \phi_0 + \frac{1}{m} \mathrm{arctan2}(\beta_m,\,\alpha_m) $. In this case, $a_m'$ is always non-negative and the `boxy/diskyness' of $m\!=\!4$ multipole depends on the range of the misalignment $\phi_4-\phi_0$, which is more tricky to recognize.
\par
\section{Optimization Details}
\label{app:optimization}
The loss function to be minimized in the optimization process is defined as the sum of two JS divergences, one for $m=3$ and the other for $m=4$:

\begin{equation}
\begin{split}
  {L}_{\mathrm{JS}} \ = \ &\mathrm{JS}_{m=3} + \mathrm{JS}_{m=4}\\
  \ =\ & \mathrm{JS}\!\bigl[P_{\mathrm{obs}}(\X3,\Y3,\Z)\,\Vert\,
                      P_{\mathrm{mod}}(\X3,\Y3,\Z)\bigr]
  \ + \ \\
  & \mathrm{JS}\!\bigl[P_{\mathrm{obs}}(\X4,\Y4,\Z)\,\Vert\,
                      P_{\mathrm{mod}}(\X4,\Y4,\Z)\bigr]\, .
\end{split}
\end{equation}
The optimization process optimized all variables shown in Table~\ref{tab:bestfit} using the \textsc{Adam} optimizer in \texttt{PyTorch}, where
\( P_{\mathrm{obs}} \) is a kernel‑density estimate of the sample from H06 \cite{hao2006isophotal}. 
The model likelihood \( P_{\mathrm{mod}} \) is evaluated by Eq.~\eqref{eq:factorization2}.
Each galaxy is represented by a
3‑D Gaussian centered at the measured value, with widths given by its quoted
uncertainties. For angular variables ($\Y3$, $\Y4$), we replicate the kernel
across the periodic boundaries\footnote{For $\Y3$ the fundamental domain is
$(-\pi/6,\pi/6)$, for $\Y4$ it is $(-\pi/8,\pi/8)$.  Any kernel that
crosses a boundary is mirrored with the appropriate parity in $\X3$ or
$\X4$; see Appendix~\ref{app:obs_kde}.}.

JS divergence is defined as
\begin{align}
\mathrm{JS}[P \Vert Q]
= \frac{1}{2} \mathrm{KL}[P \Vert M]
+ \frac{1}{2} \mathrm{KL}[Q \Vert M], \quad
\text{where } M = \frac{1}{2}(P + Q),
\end{align}
and the Kullback–Leibler divergence is given by
\begin{align}
\mathrm{KL}(P \Vert Q) = \int P(x) \log \frac{P(x)}{Q(x)} \ \d x_1\,\d x_2\,\cdots\d x_n.
\end{align}
The JS divergence provides a symmetric and smooth measure of dissimilarity between probability distributions and is widely used in the optimization of probabilistic models. JS divergence has the lower bound of $0$ when $P(x)$ and $Q(x)$ are identical and the upper bound of $\log(2)\approx0.69$ when $P(x)$ and $Q(x)$ are completely disjoint. After optimization procedure\footnote{$5000$ epochs with adaptive learning rate (ReduceLROnPlateau) starting with $0.01$ with patience 50 and factor 0.5}, it reaches to $\mathrm{JS}_{m=3}=0.11$ and $\mathrm{JS}_{m=4}=0.16$.


\section{Generation of the observational probability densities}
\label{app:obs_kde}

The \cite{hao2006isophotal} catalog provides $(\alpha_m,\ \beta_m,\ \Z)$ for $m=3$ and $m=4$ together with $1\sigma$ measurement uncertainties
$(\Delta\alpha_{m},\ \Delta\beta_{m},\ \Delta\Z)$. This is converted to $(\X{m},\Y{m},\Z)$ using the conversion shown in Appendix \ref{appendix:multipole_convention}. The uncertainties $(\Delta\X{m},\ \Delta\Y{m},\ \Delta\Z)$ are calculated using linear error analysis, but there were data points showing imaginary uncertainties from linear analysis and their uncertainties were calculated again quadratically. There were 7 out of 847 data points that still showed imaginary uncertainties even then, which were removed from this analysis.

To compare our hierarchical model with the data, we convert this discrete sample into two continuous and normalized probability densities
\(
  P_{\rm obs}(\X3,\Y3,\Z)\) and
\(P_{\rm obs}(\X4,\Y4,\Z)\).
Our procedure is summarized below.

\begin{enumerate}[label=\arabic*., leftmargin=1.2em]
\item\textbf{Gaussian kernel.}  
      Each galaxy $i$ is represented by a 3‑D Gaussian
\[
  G_i = 
  \mathcal{N}\!\bigl((\X{m})_i,\Delta{(\X{m})_i}^2\bigr)\;
  \mathcal{N}\!\bigl((\Y{m})_i,\Delta{(\Y{m})_i}^2\bigr)\;
  \mathcal{N}\!\bigl({\Z}_i ,\Delta{\Z_i}^2\bigr).
\]
\item\textbf{Angular wrapping.}  
      Because the angular variables are periodic,
      $\Y{m}\in\bigl(-\tfrac{\pi}{2m},\tfrac{\pi}{2m}\bigr)$,
      a kernel whose width extends beyond the boundary must be
      replicated.  
      We compute
      \(
        n = \bigl\lceil
             \lvert (\Y{m})_i \pm 3\Delta(\Y{m})_i\rvert\,
             \big/(\pi/2m)
            \bigr\rceil
      \)
      and create shifted copies
      \(
        ({(\X{m})_i}',{(\Y{m})_i}',{\Z_i}') =
        (\pm(\X{m})_i,\,(\Y{m})_i\pm 2k\!\cdot\!\pi/2m,\,{\Z}_i)
      \)
      for every $k=-n,\dots,n$.  
      Odd–$k$ shifts flip the sign of $\X{m}$ because the $m$‑th
      multipole is antisymmetric under a half‑period rotation.
\item\textbf{Weighting.}  
      Each replica is assigned the analytical weight
      \(
        w_k =
        \mathrm{CDF}_{\mathcal{N}}
        \bigl(\tfrac{\pi}{2m}+2k\tfrac{\pi}{2m}\bigr)
        -
        \mathrm{CDF}_{\mathcal{N}}
        \bigl(-\tfrac{\pi}{2m}+2k\tfrac{\pi}{2m}\bigr).
      \)
      This ensures that the integral of the replicated kernels equals
      unity.  All kernels share the same weight in $\Y{m}$ and $\Z$.
\item\textbf{Normalization.}  
      The sum
      \(
        P_{\rm obs} = \sum_i \sum_k w_{ik}\,G_{ik}
      \)
      is evaluated on a regular $(100)^3$ grid and renormalised so that
      $\int P_{\rm obs}=1$.
\end{enumerate}

\section{Modeling details}
\label{appendix:modeling_details}

Table \ref{table:lens_mass_params} and \ref{table:light_params} 
provide the lens and light parameters other than the multipole parameters used to create the mock data.

\begin{table*}[h]
\centering
\begin{adjustbox}{center}
\begin{tabular}{c c c l c}
Profile & \makecell{Parameter\\Name} & True Value  &Prior& Note \\
\Xhline{3\arrayrulewidth}
\multirow{8}{*}{\makecell{Elliptical\\Power Law\\(EPL)}} 
 & $\theta_E$ & $1\farcs37$  & $U(0,10)$ &  \\
 & $\gamma$ & 2.50  &  $U(1.5, 2.5)$ &  \\
 & $x_\mathrm{center}$& $0\farcs043$  &  $U(-10, 10)$  &  \\
 & $y_\mathrm{center}$& $0\farcs002$  &  $U(-10, 10)$  &  \\
 & $e_1$& $-0.07$&  \multirow{2}{*}{\makecell{See Appendix \ref{app:e1e2_prior}}}  &   \\
 & $e_2$& $-0.25$&    &  \\
 &\cellcolor{LightGray} $q$ &\cellcolor{LightGray} $0.58$ &\cellcolor{LightGray} Converted from $(e_1, e_2)$ & (a)
 \\
 &\cellcolor{LightGray} $\phi_0$ &\cellcolor{LightGray} $-0.93 \ (-53^\circ)$ &\cellcolor{LightGray} Converted from $(e_1, e_2)$ & (b) \\ 
 \hline
\multirow{4}{*}{External Shear} 
& $\gamma_1$ & 0.04 & $U(-0.5,0.5)$ &  
\\
& $\gamma_2$ & 0.10 &  $U(-0.5,0.5)$  &  
\\
 &\cellcolor{LightGray} $\gamma_{\mathrm{ext}}$ &\cellcolor{LightGray} $0.10$  &\cellcolor{LightGray} Converted from $(\gamma_1, \gamma_2)$ & (c) \\
&\cellcolor{LightGray} $\phi_{\mathrm{ext}}$ &\cellcolor{LightGray} $0.60$ ($34^\circ$)  &\cellcolor{LightGray} Converted from $(\gamma_1, \gamma_2)$ & (d) \\
 \hline
\multirow{4}{*}{\makecell{$m=3$ Elliptical\\Multipole}} 
& $a_3\!/\!a$ & \multicolumn{2}{c}{\multirow{2}{*}{See Table \ref{table:m=34_true_params}}}  &  
\\
 & $\phi_3-\phi_0$ &   &  & \\
 & \cellcolor{LightGray} $x_\mathrm{center}$ & \cellcolor{LightGray} $0\farcs043$ & \cellcolor{LightGray} Jointly sampled with \\
 &\cellcolor{LightGray} $y_\mathrm{center}$ & \cellcolor{LightGray} $0\farcs002$ & \cellcolor{LightGray} EPL's $(x_\mathrm{center}, y_\mathrm{center})$ \\
 \hline
\multirow{4}{*}{\makecell{$m=4$ Elliptical\\Multipole}}
   & $a_4\!/\!a$ & \multicolumn{2}{c}{\multirow{2}{*}{See Table \ref{table:m=34_true_params}}}  & \\
   & $\phi_4$ &  & \\
 & \cellcolor{LightGray} $x_\mathrm{center}$ & \cellcolor{LightGray} $0\farcs043$ & \cellcolor{LightGray} Jointly sampled with \\
 &\cellcolor{LightGray} $y_\mathrm{center}$ & \cellcolor{LightGray} $0\farcs002$ & \cellcolor{LightGray} EPL's $(x_\mathrm{center}, y_\mathrm{center})$ \\
    \hline
\end{tabular}
\end{adjustbox}
    \caption{
    The true values and priors of the lens mass parameters used for simulation and fitting of the lensed quasar system. \\(a)
    $q = \frac{1-c}{1+c},\ c=\sqrt{{e_1}^2+{e_2}^2}$.
    (b) $\phi_0 = \frac{1}{2} \arctan\!2(e_2, e_1)$.
    (c) $\gamma_\mathrm{ext} = \sqrt{{\gamma_1}^2+{\gamma_2}^2} $. 
    (d) $\phi_\mathrm{ext}=\frac{1}{2} \arctan\!2(\gamma_2, \gamma_1) $.
    }
    \label{table:lens_mass_params}
\end{table*}

\begin{table}[h]
    \centering
\begin{adjustbox}{center}
\begin{tabular}{c c c l c}
Kind & \makecell{Parameter\\Name} & True Value  &Prior& Note \\
\Xhline{3\arrayrulewidth}
\multirow{2}{*}{Quasar} 
 & $x_\mathrm{source}$& $0\farcs18$ & Not directly sampled & \multirow{2}{*}{\makecell{(a)}} \\ 
 & $y_\mathrm{source}$& $-0\farcs10$ & Not directly sampled & \\
 \hline
\multirow{9}{*}{\makecell{Elliptical Sérsic\\(Source Light,\\when arcs exist)}}& $I_e$ & 40  &Not directly sampled& \\
 & $R_\mathrm{sersic}$ & $0\farcs37$&$U(0.001,10)$& \\
 & $n_\mathrm{sersic}$ & $1.0$  &$U(0.5,5)$&  \\
 & \cellcolor{LightGray} $x_\mathrm{source}$ & \cellcolor{LightGray} $0\farcs18$ & \cellcolor{LightGray} Jointly sampled with &  \\
 & \cellcolor{LightGray} $y_\mathrm{source}$ & \cellcolor{LightGray} $-0\farcs10$ & \cellcolor{LightGray} Quasar's $(x_\mathrm{source}, y_\mathrm{source})$ &  \\
 & $e_1$& $0.37$  &$U(-0.5,0.5)$&  \\
 & $e_2$& $0.13$  &$U(-0.5,0.5)$&  \\
 &\cellcolor{LightGray} $q_\text{source}$ &\cellcolor{LightGray}0.43 &\cellcolor{LightGray} Converted from $(e_1, e_2)$ & 
    (b)
 \\
 &\cellcolor{LightGray} $\phi_\text{source}$ &\cellcolor{LightGray} $0.17\ (9.7^\circ)$ &\cellcolor{LightGray} Converted from $(e_1, e_2)$ & (c) \\ 
 \hline
\multirow{9}{*}{\makecell{Elliptical Sérsic\\(Lens Light)}}
& $I_e$ & 12  & Not directly sampled &  \\
 & $R_\mathrm{sersic}$ & $3.3$ & $U(0.001, 10)$ &  \\
 & $n_\mathrm{sersic}$ & $3.9$  & $U(0.5, 5)$ &  \\
 & $x_\text{lens}$ & $0\farcs03$& $U(-10,10)$ &  \\
 & $y_\text{lens}$ & $0\farcs01$& $U(-10,10)$ &  \\
 & $e_1$ & $-0.05$  & $U(-0.5,0.5)$ &  \\
 & $e_2$ & $-0.18$  & $U(-0.5,0.5)$ &  \\
 &\cellcolor{LightGray} $q_\mathrm{lens}$ &\cellcolor{LightGray} $0.69$ &\cellcolor{LightGray} Converted from $(e_1, e_2)$ & 
    (b)
 \\
 &\cellcolor{LightGray} $\phi_\mathrm{lens}$ &\cellcolor{LightGray} $-0.92 \ (-52^\circ)$ &\cellcolor{LightGray} Converted from $(e_1, e_2)$ & (c) \\ 
\hline
\end{tabular}
\end{adjustbox}
\caption{The true values and priors of the source and lens light parameters used for simulation and fitting of the lensed quasar system.
\\
(a) The lensed positions are sampled first and their unlensed position was evaluated.
(b) $q = \frac{1-c}{1+c}, \  c=\sqrt{{e_1}^2+{e_2}^2}$.
(c) $\phi = \frac{1}{2} \arctan\!2(e_2, e_1)$
}
    \label{table:light_params}
\end{table}

\section{Ellipticity parameterization and prior}
\label{app:e1e2_prior}

The multipole prior \(
P(\X{3},\ \Y{3},\ \X{4},\ \Y{4},\ \Z)
\) discussed in Section~\ref{sec:prior}
provides a skew-normal distribution as the prior for $\Z$, and this is implemented for the prior of $(e_1,\,e_2)$, considering their relationship with $\Z$, which is $q = \frac{1-c}{1+c},\ c=\sqrt{{e_1}^2+{e_2}^2}$. However, this is not enough because the parameterization of $(e_1,\,e_2)$ itself carries a strong bias toward small $\Z$ and thus affect the multipole prior too, if not adjusted. In the following content, we explain this problem and an additional prior to relieve this bias.

\begin{figure}[h]
  \centering
  \includegraphics[width=0.95\textwidth]
    {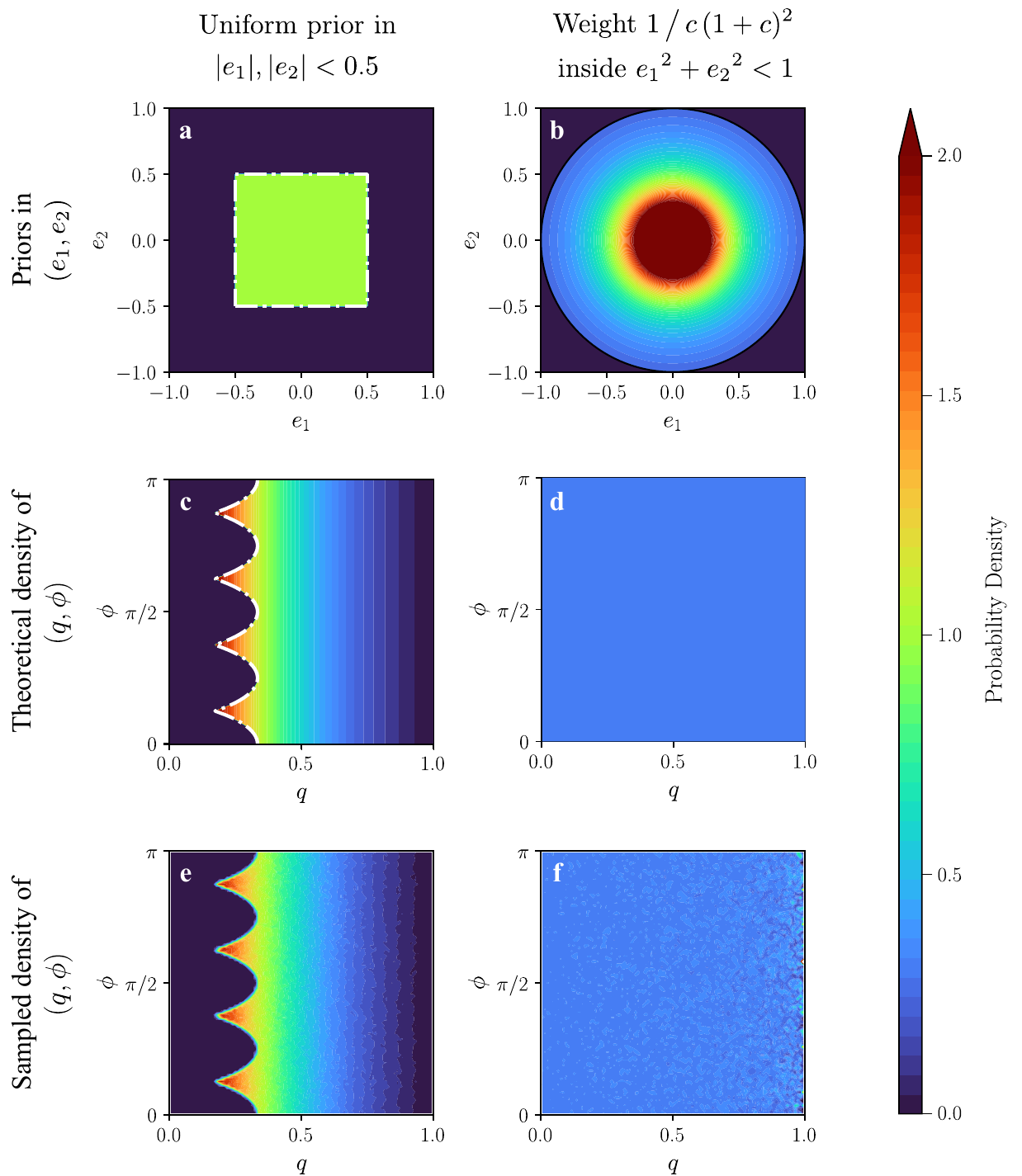}
  \caption{\textbf{Top:} Priors in the $(e_1, e_2)$ space.
\textbf{(a)} Uniform in the square $|e_1|, |e_2| \le 0.5$.
\textbf{(b)} Weighted in the unit disk ($c<1$) by $1 \,/\, c\, (1+c)^2 $.
\\
\textbf{Middle:} Analytic mapping to $(q, \phi)$.
\textbf{(c)}
Flat square in $(e_1, e_2)$ maps to a biased $(q, \phi)$ prior, increasing sharply as $q \to 0$, with a hard cutoff at $q > 3{-}2\sqrt{2}$. The white dash-dot line traces the image of the square boundary from panel (a).
\textbf{(d)}
Applying the weight $\frac{1}{c\,(1+c)^2}$ flattens the prior over $(q,\phi)$.
\\
\textbf{Bottom:} Monte Carlo validation.
\textbf{(e)} Sampling from the square reproduces (c).
\textbf{(f)} Disk sampling with weight recovers uniformity as in (d). The granular structure at high $q$ results partly from limited number of sampling ($N=5{\times}10^6$). Additionally, near $q \sim 1$ (i.e., $e_1, e_2 \sim 0$), numerical precision limits in $(e_1, e_2)$ introduce artifacts in the transformed $(q, \phi)$ space.}
  \label{fig:e1e2_prior_demo}
\end{figure}

\paragraph{Definitions.}
The axis ratio $q\!=\!b/a\,(0<q\le1)$ and major‑axis
position angle $\phi\!\in[0,\pi)$ are mapped to the
Cartesian ellipticity components%
\footnote{This is the \textsc{lens\-tronomy} convention.}
\begin{equation}
  \label{eq:e1e2_def}
  c   \;=\; \frac{1-q}{1+q},\qquad
  e_1 \;=\; c\cos 2\phi,\qquad
  e_2 \;=\; c\sin 2\phi.
\end{equation}
Sampling in the $(e_1,e_2)$ avoids the discontinuity at
$\phi\!=\!0=\!\pi$ and helps a continuous angular sampling, but introduces a \emph{non‑uniform} Jacobian with
respect to $(q,\phi)$:
\begin{equation}
  \label{eq:jacobian}
  \det \left(
  \frac{\partial(e_1,e_2)}{\partial(q, \phi)}
  \right)
  \;=\;
  \left| \begin{matrix}
    \frac{\partial e_1}{\partial q} & \frac{\partial e_1}{\partial \phi} \\
    \frac{\partial e_2}{\partial q} & \frac{\partial e_2}{\partial \phi}
    \end{matrix}
    \right |
    \;=\; 
  -\,\frac{4 (1-q)}{(1+q)^3} 
  \;=\;
  - c\, (1+c)^2.
\end{equation}
A flat prior in $(e_1,e_2)$ therefore induces a strong preference for
smaller ellipticities ($q\!<\!0.5$) and cannot reach $q\!<\!3-2\sqrt{2}\approx0.17$ with any angle and $3-2\sqrt{2} \le q\!<\!1/3$ for some angles, if the usual box $|e_1|,|e_2|\le0.5$ is enforced.
Figure~\ref{fig:e1e2_prior_demo} (a, c, e) illustrates the bias.
\paragraph{Desired prior.}
For our lens population we wish to keep
$\!P(q)\!\propto\!\mathrm{const}$ and
$P(\phi)\!\propto\!\mathrm{const}$ before giving the multipole-related prior.
Combining Eq.~\eqref{eq:jacobian} with
$P(q,\phi)=P(e_1,e_2)\,|\det J|$ yields
\begin{equation}
  P(e_1,e_2)\;\propto\;
  \frac{(1+q)^3}{4 (1-q)}\;
  =\;\frac{1}{c(1+c)^2}
  \label{eq:e_prior}
\end{equation}
The prior applies the entire unit disk
${e_1}^2+{e_2}^2<1$, allowing axis ratios down to $q=0$.
We implement this as an additive custom log‑prior in the
\textsc{lenstronomy} likelihood:

\begin{align}
    \ln\mathcal{L}_{e}
  \;&=\;
  -\ln\!\bigl[c\,(1+c)^2\bigr]\;\; & \text{if } {e_1}^2+{e_2}^2<1,
  \\
  \ln\mathcal{L}_{e}
  \;&=\;
  -\infty \;\; & \text{otherwise}.
  \label{eq:log_prior}
\end{align}
Figure~\ref{fig:e1e2_prior_demo} (b, d, f) illustrates the sampling result with the desired prior.

\vspace{1 cm}

\bibliographystyle{JHEP}
\bibliography{references.bib}

\end{document}